\documentclass{aa}
\pdfoutput=1
\usepackage{amsmath}
\usepackage[varg]{txfonts}
\usepackage{graphicx}
\usepackage{natbib}
\usepackage{amssymb}
\usepackage{color}

\bibpunct{(}{)}{;}{a}{}{,} 

\newcommand{\HALF}{\frac{1}{2}}
\newcommand{\pd}[2]{\frac{\partial #1}{\partial #2}}
\newcommand{\DS}{\displaystyle}
\renewcommand{\vec}[1]{\mathbf{#1}}
\newcommand{\hvec}[1]{\hat{\mathbf{#1}}}

\begin{document}
\title{Astrophysical fluid simulations of thermally ideal gases with non-constant adiabatic index: numerical implementation}

\subtitle{}

\author{B. Vaidya\inst{\ref{inst1}} \and A. Mignone\inst{\ref{inst1}} \and G. Bodo\inst{\ref{inst2}} \and S. Massaglia\inst{\ref{inst1}}}
\institute{Dipartimento di Fisica, Università di Torino, via Pietro
  Giuria 1, I-10125 Torino, Italy \label{inst1} \and INAF, Osservatorio Astronomico di Torino, Strada Osservatorio 20, I-10025 Pino Torinese, Italy \label{inst2}}

\date{Received date / Accepted date}

\abstract
{
An Equation of State (\textit{EoS}) is a relation between thermodynamic state variables and it is essential for closing the set of equations describing a fluid system.
Although an ideal EoS with a constant \textit{adiabatic index} $\Gamma$ is the preferred choice due to its simplistic implementation, many astrophysical fluid simulations may benefit from a more sophisticated treatment that can account for diverse chemical processes.
}
{
In the present work we first review the basic thermodynamic principles
of a gas mixture in terms of its thermal and caloric EoS by including
effects like ionization, dissociation as well as temperature dependent degrees of freedom such as molecular vibrations and rotations.
The formulation is revisited in the context of plasmas that are either in equilibrium conditions (local thermodynamic- or collisional excitation- equilibria) or described by non-equilibrium chemistry coupled to optically thin radiative cooling.

We then present a numerical implementation of thermally ideal gases obeying a more general caloric EoS with non-constant adiabatic index in Godunov-type numerical schemes.
}
{
We discuss the necessary modifications to the Riemann solver and to the conversion between total energy and pressure (or vice-versa) routinely invoked in Godunov-type schemes. 
We then present two different approaches for computing the EoS.
The first one employs root-finder methods and it is best suited for EoS in analytical form.
The second one leans on lookup table and interpolation and results in a more computationally efficient approach although care must be taken to ensure thermodynamic consistency.
}
{
A number of selected benchmarks demonstrate that the employment of a non-ideal EoS can lead to important differences in the solution when the temperature range is $500-10^4$  K where dissociation and ionization occur. 
The implementation of selected EoS introduces additional computational costs although the employment of lookup table methods (when possible) can significantly reduce the overhead by a factor $3\sim 4$.
}
{}


\titlerunning{Caloric EOS in {\it{PLUTO}}}
\authorrunning{Vaidya et al.}

\maketitle
\section{Introduction}
%
%
%
%

An equation of state (EoS) is a relationship between state variables of a thermodynamic system under certain physical conditions.
Such a constitutive equation provide a necessary closure for a complete mathematical description of a fluid system in addition to the conservation laws of mass, momentum and energy.
Numerical simulations of astrophysical systems such as inter-stellar medium, planetary atmospheres, stellar evolution,  jets and outflows, require inter-play of various thermal, radiative and chemical processes.
For such complex systems, using a simple ideal (or an isothermal) EoS would be considered as a serious limitation.
A consistent description for such systems demands the use of a \textit{general EoS} that can account for thermal and chemical processes.

For example, the thermodynamic state of the gas plays a pivotal role in governing the fragmentation of self-gravitating and turbulent molecular clouds \citep[e.g.,][]{Spaans2000, Li2003, Jappsen2005}.
The balance between heating and cooling in molecular clouds is approximated by using a poly-tropic EoS, $p\propto \rho^{\Gamma}$.
Multiple smoothed particle hydrodynamical simulations with different adiabatic indices, $0.2 < \Gamma <  1.4$ (\citealt{Spaans2000}) were used to show that the degree of fragmentation decreases with increasing value of $\Gamma$ (\citealt{Li2003}).
\cite{Jappsen2005} showed that the thermal properties of the gas determines the stellar mass function (IMF) using a piecewise poly-tropic EoS.
Such empirical forms of EoS in general depend on chemical abundances and complex atomic and molecular physics. 

Numerical simulations studying thermo-chemical evolution of  early structure formation used an \textit{effective} adiabatic index, $\Gamma_{\rm eff}$, to relate internal energy with thermal pressure \citep[e.g.][]{Yoshida2006, Glover2008}.
The value of $\Gamma_{\rm eff}$ is estimated from number fractions of chemical species treating the chemical composition as an ideal mixture. 
In the context of disk instability leading to formation of gas giant planets, \cite{Boley2007} pointed out the importance of incorporating isotopic forms of molecular hydrogen, $H_{2}$, as well the molecular physics (rotation and vibration) under thermodynamic equilibrium in the estimate of internal energy.
A more complex EoS taking into account ionization of atomic hydrogen, helium and radiation along with molecular dissociations is used to study the envelopes of young planetary cores \citep{DAngelo2013}.  

From the computational perspective, procedures to treat
  dynamics of astrophysical plasmas with a general
EoS have been developed by appropriately modifying the Riemann solver
as described, e.g., by \cite{Colella1985, Glaister1988a, Glaister1988b,
  Menikoff:1989aa, Liou1990, Fedkiw1997, Hu2009}. Similarly, 
numerical codes like FLASH \citep{Fryxell2000} and CASTRO
\citep{Almgren2010} have implemented electron-positron
EoS \citep{Timmes2000} using high order polynomials as interpolating functions.
Such an EoS based on tabulated Helmholtz free energy is employed in
the study of stellar evolution and supernovae. 
\cite{Falle1993, Falle1995} incorporated LTE effects to model
partially ionized hydrogen gas using Saha Equations to 
study variable knots produced in stellar jets.

The goal of this paper is to outline a consistent numerical framework
for the implementation of a more general equation of state in the
context of the magnetohydrodynamics (MHD) equations. 
Our formulation accounts for different physical processes such as atomic ionization and recombination, molecular dissociation, etc... and it is suitable under equilibrium conditions (local thermodynamic or collisional ionization equilibria) as well as for non-equilibrium optically thin radiative cooling \citep{Tesileanu2008}.
The numerical method is implemented as part of the PLUTO code \citep{Mignone2007} and it is built while ensuring thermodynamical consistency, accuracy and computational efficiency.

Our starting point are the ideal MHD equations written in conservation form:
\begin{eqnarray}\label{eq:mhd_rho}
\DS \pd{\rho}{t} + \nabla\cdot\left(\rho\vec{v}\right) &=& 0 
 \\ \noalign{\medskip} \label{eq:mhd_mom}
\DS \pd{\vec{(\rho\vec{v})}}{t} + \nabla\cdot\left(
   \rho\vec{v}\vec{v}^T - \vec{B}\vec{B}^T\right) + \nabla p_t&=& 0
\\ \noalign{\medskip}\label{eq:mhd_B}
\DS \pd{\vec{B}}{t} - \nabla\times\left(\vec{v}\times\vec{B}\right) 
   &=& \vec{0}\,
\\ \noalign{\medskip}\label{eq:mhd_E}
\DS \pd{E}{t} + \nabla\cdot\left[
  \left(E + p_t\right)\vec{v} - 
  \left(\vec{v}\cdot\vec{B}\right)\vec{B}\right]& = & \Lambda
\\ \noalign{\medskip}\label{eq:mhd_X}
\DS \pd{(\rho X_k)}{t} + \nabla\cdot\left(\rho X_k\vec{v}\right) &=& S_k
\end{eqnarray}
where $\rho$ is the mass density, $\vec{v}$ is the fluid velocity, $\vec{B}$ is the magnetic field, $p_t=p+\vec{B}^2/2$ is the total pressure accounting for thermal ($p$) and magnetic ($\vec{B}^2/2$) contributions. 
The total energy density $E$ is given by
\begin{equation}\label{eq:total_energy}
 E = \rho e + \frac{1}{2}\rho\vec{v}^2
            + \frac{1}{2}\vec{B}^2\,.
\end{equation}
An additional EoS relating the internal energy density $\rho e$ with $p$ and $\rho$ must be specified.
This issue is addressed in \S \ref{sec:eos}.
Dissipative effects have been neglected for the sake of exposition although they can be easily incorporated in this framework.

The paper is organized as follows, in \S\ref {sec:eos} the basic principles and formulations of general \textit{EoS} used for the present work are described.
The numerical framework is discussed in \S\ref{sec:numframe}.
The results obtained from various test problems are outlined in \S \ref{sec:tests} and the concluding remarks are summarized in \S \ref{sec:summary} . 

\section{Equation of State}
\label{sec:eos}
%
%
%

\subsection{Thermodynamical Principles}
%
%

The principle of conservation of energy in thermodynamics is commonly known as the
first law of thermodynamics and can be expressed as,
\begin{equation} \label{eq:2}
d\textit{U}  = T d\textit{S}  - p d\textit{V}\,,
\end{equation}
where $S(U, V)$ is the entropy.
The internal energy $U$ and the volume $V$ are classified as
extensive variables and depend on bulk properties of the
system. Whereas, the intensive variables like temperature $T$ and
pressure $p$ show no dependence on the size of the system.
An \textit{EoS} describing such a system is defined as a relation among intensive and extensive variables.
A {\em thermal} EoS is an expression relating pressure, temperature and volume and we will express it as $p = p (V,T)$.
Conversely, a {\em caloric} EoS specifies the dependence of the internal energy of the system $U$ on volume $V$ and temperature $T$.
The total internal energy, $U$ is related to the internal energy density (see Eq. \ref{eq:total_energy}) as $U/V = \rho e$.

In general, different forms of EoS relations are derived from empirical results and are used to estimate various thermodynamical properties of a system.
Theoretically, statistical principles can be applied to describe such a system on basis of its microscopic processes using the partition function $\mathcal{Z}$.
For example, the macroscopic thermodynamic quantities can be obtained from the following standard relations, 
\begin{equation}\label{eq:3}
\begin{array}{l}
  \DS p = k_BT \left(\frac{\partial\ln \mathcal{Z}}{\partial V}\right)_{T}\\
\\
 \DS U = k_BT^{2}\left(\frac{\partial \ln \mathcal{Z}}
                                {\partial T}\right)_{V},
\end{array}
\end{equation}
where, $k_B$ is the Boltzmann constant.
The previous equation essentially provides two forms of \textit{EoS} in terms of partition function. 

An important quantity that relates the pressure $p$ with density $\rho$ is the speed of sound, defined as 
\begin{equation}\label{eq:cs}
c_s = \sqrt{\left(\frac{\partial p}{\partial \rho}\right)_{s}}.
\end{equation}
where the derivative must be taken at constant entropy.
The above definition can be further expressed in terms of the first adiabatic exponent, $\Gamma_1$, defined as ($\partial \ln p/\partial \ln \rho$)$_{s}$.
Thus, Eq. (\ref{eq:cs}) now becomes
\begin{equation}\label{eq:cs2}
  c_s = \sqrt{\left(\frac{\Gamma_1 p}{ \rho}\right)}.
\end{equation}
In general the first adiabatic exponent $\Gamma_1$ has a functional dependence on temperature and density as,
\begin{equation}
\label{eq:gam1}
\Gamma_1 = \frac{1}{C_{V}(T)} \left(\frac{p}{\rho T}\right)\chi_T^{2}
+ \chi_\rho,
\end{equation}
where, $C_V(T)$ is obtained by taking the derivative of the specific
gas internal energy, $e(T)$ with respect to temperature at constant
volume while $\chi_T$ and $\chi_P$ are
referred to as temperature and density exponents (see
\cite{DAngelo2013})
\begin{equation}\label{eq:temprhoexp}
\begin{array}{l}
\DS \chi_T = \left(\frac{\partial\, \rm{ln}\, p}{\partial \,\rm{ln}\, T}\right)_{\rho} = 1 -
  \frac{\partial \,\rm{ln}\, \mu}{\partial\, \rm{ln}\, T}\\
\\
\DS \chi_\rho = \left(\frac{\partial\, \rm{ln}\, p}{\partial\, \rm{ln}\, \rho}\right)_{T} = 1 -
  \frac{\partial\, \rm{ln} \, \mu}{\partial \, \rm{ln} \, \rho}.
\end{array}
\end{equation}

For an ideal gas, the value of $\Gamma_1$ coincides with the adiabatic index $\Gamma$, which is essentially the ratio of specific heats. 
In such a case, the sound speed can be expressed as
\begin{equation}\label{eq:cs3}
c_s = \sqrt{\left(\frac{\Gamma p}{\rho}\right)}\,.
\end{equation}

In the present work, we will focus on {\em thermally ideal gases}.
These gases have their thermal EoS same as that of an ideal gas.
However, the caloric EoS can have nonlinear dependence on temperature based on various chemical processes taken into consideration (see \S\ref{thermideal}). 
We point out that, although the analysis presented here is limited to thermally ideal gas, the numerical implementation described in this work can also be extended to describe real gases obeying EoS that are not thermally ideal. 

\subsection{Thermodynamic Constraints}
\label{sec:thermodynamic_constraints}
%
%

The equation of state must adhere to a number of physical principles in order to be thermodynamically consistent.
Although a comprehensive discussion lies outside the scope of this
paper, we briefly outline the most relevant ones \citep{Menikoff:1989aa}.

\begin{enumerate}
  \item The specific internal energy as a function of specific volume and
        entropy $e=e(V,S)$ must be piecewise twice continuously differentiable.
  \item Thermodynamic stability demands that $e(V,S)$ be a jointly
        convex function.
        This implies that the Hessian matrix of second derivatives of $e$ with
        respect to $V$ and $S$ is non-negative.
  \item Simple physical considerations lead to the following asymptotic conditions:
        \begin{eqnarray}
             \lim_{V\to\infty} p(V,S) &=&  0  \\ \noalign{\medskip}
             \lim_{V\to 0}     p(V,S) &=&  \lim_{S\to\infty} p(V,S)
           = \lim_{S\to\infty} e(V,s) = \infty
        \end{eqnarray}
        The previous constraints also guarantee the existence of
        the solution of the Riemann problem.
\end{enumerate}

An additional constraint, convexity, can be introduced if one sticks to standard theory and phase transition are not considered.
The convexity of an EoS is quantified by the {\em{fundamental gas derivative}} which expresses the nonlinear variation of the sound speed with respect to density and it is denoted by $\mathcal{G}$:
\begin{equation}\label{eq:G}
\mathcal{G} = 1 + \frac{\rho}{c_s}\left(\frac{\partial c_s}
                                             {\partial\rho}\right)_{s}\,.
\end{equation}
In Eq. (\ref{eq:G}) the derivative is taken at constant entropy and $c_s$ is the speed of sound given by Eq. (\ref{eq:cs}). 
For an ideal polytropic gas, one finds $\mathcal{G} = (\Gamma + 1)/2$ while a general expression in terms of derivatives with respect to temperature and density may be found in Appendix \ref{sec:G}.

An EoS is said to be convex if $\mathcal{G} > 0$ and it has the following important implications: i) the isoentropes are convex functions in the $p-V$ plane, ii) the sound speed increases with density along isoentropes, iii) only regular waves (e.g. compression shock waves and expansion fans) can be formed in the Riemann problem.

The latter property is of particular interest here since, as we shall see, inaccurate table interpolation can lead to local violation of the convexity assumption (see the discussion in Section \ref{sec:interpolation} and the results in Appendix \ref{sec:Gtest}).
In such cases, composite (or compound) waves consisting of a rarefaction wave propagating adjacent to a shock may be generated in the solution whilst satisfying the thermodynamical principles \citep[e.g.][and references therein]{Menikoff:1989aa}.
This circumstance may arise, for instance, for a real gas in correspondence of finite intervals of {\em{concave}} $p-V$ isoentropes.

In the present paper, however, we restrict our attention to EoS for which $\mathcal{G}>0$ is always verified at the continuous level although it may not be true at the discrete numerical level thereby generating spurious composite waves.
We address this issue in Sect. \ref{sec:interpolation}.

\subsection{Calorically Ideal Gas}
\label{thermideal}
%
%

Consider the case of a classical monoatomic ideal gas, where, the partition function $\mathcal{Z}$ is given by
\begin{equation}\label{eq:Zideal}
\mathcal{Z} = \frac{1}{N!}\left[\left( \frac{mk_BT}{2\pi\hbar^2}\right )^{3/2}\textit{V}\right]^{N},
\end{equation}
where, $m$ is the mass of the particle, $\hbar$ the Planck constant and $N$ the total number of non-interacting particles.
On substituting Eq. (\ref{eq:Zideal}) in Eq. (\ref{eq:3}), we obtain the standard
EoS for a classical ideal gas,
\begin{equation}\label{eq:ideal}
\begin{array}{l}
  p = nk_BT\\    \\
  \DS \rho\textit{e} = U/V = \frac{f_{\rm trans}}{2}nk_BT,
\end{array}
\end{equation}
where,   $n = N/V$ is the number density and $f_{\rm trans}$ denotes
the translational degree of freedom which for a monoatomic gas equals
to 3.
Further, the specific heat capacity at constant volume, $C_V = f_{\rm trans}R/2$ (where $R$ being the universal gas constant), is independent of the temperature.
On extending this analysis further to diatomic ideal gas, the partition function $\mathcal{Z}$ contains contribution from rotational and vibrational degrees of freedom, in addition to the translational motion.
In such a case, the internal energy density derived from Eq.\ref{eq:3} is given by,
\begin{equation} \label{eq:5a}
\rho \textit{e} = \frac{f_{\rm trans}}{2}nk_BT + \frac{f_{\rm rot}}{2} nk_BT + \Phi_{\rm vib}(T)
\end{equation}
where the additional contribution of $f_{\rm rot}nk_BT/2$ comes from $f_{\rm rot}$  rotational degree of freedoms, whose value is 2 for linear molecules and 3 for non-linear ones. 
In addition, $\Phi_{\rm vib}(T)$ denotes term due to vibrational motion which has a non-linear dependence on temperature.
On considering the diatomic molecule with two degrees of freedom (i.e, translational and rotational) and neglecting the non-linear dependence due to vibration, one obtains a single relation for both monoatomic and diatomic gas by adopting a 
constant $\Gamma$,
\begin{equation}\label{eq:6}
p = (\Gamma - 1) \rho \textit{e},
\end{equation}
where $\Gamma = 5/3$ for monoatomic gas and $\Gamma=7/5$ for diatomic gas (see Eqns. \ref{eq:ideal} and \ref{eq:5a}).

\subsection{Partially Ionized Hydrogen Gas}
%
%

Astrophysical fluids and processes are more complex than the simple system of ideal gas described above.
For example, the Inter-Stellar Medium (ISM) that largely comprises hydrogen and helium is affected by many physical and chemical processes viz., collisional ionization, dissociation, shocks, radiation, etc.
In such a scenario, an heuristic approach that models the ISM as an monoatomic ideal gas with constant $\Gamma = 5/3$ will only be approximate and fail to account for the feedback of the above processes on the thermal properties of the gas and thereby also on its inter-linked dynamics. 

Consider the simplest case of a partially ionized gas of pure hydrogen (in atomic form).
The thermodynamics of such a system is different from that of a completely ionized (or completely neutral) gas as the number of free particles can change and an  additional energy contribution is required during the process of ionization.
The internal energy density is therefore given by (\citealt{ClaytonBook}), 
\begin{equation} \label{eq:H+EoS}
\rho e = \frac{3}{2}nk_BT + \chi_{H} n_{HII} \,.
\end{equation}
In addition to the standard form of translational energy, contribution from ionization potential, $\chi_{H}$, is included in Eq. (\ref{eq:H+EoS}).
Here, n$_{HII}$ is the number density of ionized hydrogens and the total number density of free particles, $n = n_H + 2n_{HII}$, is the sum of number densities of neutral hydrogens and twice that of $n_{HII}$ due to charge neutrality.

In regions of dense stellar interior, one can assume local thermodynamic  equilibrium.
For such a system, the fractions of ionized hydrogen becomes nonlinearly dependent on temperature and density of the gas through the Saha equation.
As a result, internal energy, specific heats and the adiabatic index $\Gamma$  will depend on the ionization fraction.
For example, the adiabatic index $\Gamma$ will smoothly change from its monoatomic value of $5/3$ to $1.13$ typical of a hydrogen gas with an ionization fraction of $50\%$ at $T = 10^4$ K (\citealt{ClaytonBook}).
Such a significant change in $\Gamma$ owes to the fact that part of the energy input becomes available to ionization rather than increasing the temperature of the gas. Therefore, using a constant value of $\Gamma$ for such dense stellar interiors will considerably overestimate the temperature of the gas. 

\subsection{Hydrogen/Helium gas mixture}
\label{sec:pvte}
%

\begin{figure*}[!ht]
\includegraphics*[width=2.0\columnwidth]{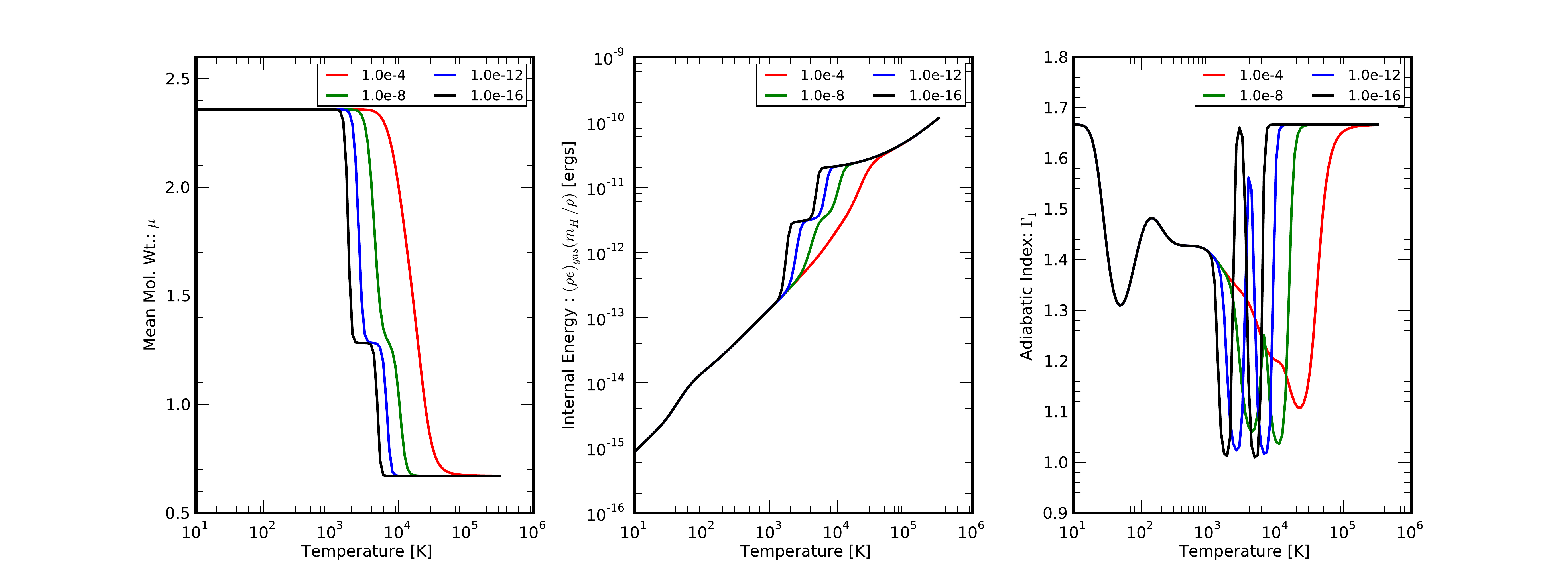}
\caption{\footnotesize Variation of Mean molecular weight $\mu$, internal energy density
of the gas $(\rho e)_{\rm gas}$ and first adiabatic index
$\Gamma_1$ with temperature. The different colored curves represent
four values of fixed density in g\,cm$^{-3}$, viz.,
10$^{-4}$ (\textit{red}),10$^{-8}$ (\textit{green}), 10$^{-12}$
(\textit{blue}) and 10$^{-16}$(\textit{black}).
The  values of $(\rho e)_{gas}$ and $\Gamma_1$ are obtained at
equilibrium between ortho and para hydrogen.}
\label{fig:muintegmm}
\end{figure*}

In recent years, studies related to planet formation in
accretion disks have started to incorporate \textit{EoS} that can account for
contributions from dissociation of molecular hydrogen, ionization of
atomic hydrogen and helium and radiation \cite[e.g.,][]{Boley2007,
  DAngelo2013} under an assumption of local thermodynamic
equilibrium (LTE). In this paper, we have implemented such an 
\textit{EoS} both in presence of LTE\footnote{For the present work, we have not considered
  contributions from radiation and $He$ ionization but rather focused on
  hydrogen alone to be consistent with the implementation in presence of cooling which assumes a fixed fraction of helium.} and with explicit non-equilibrium
cooling. For all future references to this \textit{EoS}, we will use {\em{H/He EoS}}.

In LTE, processes like ionization-recombination and
dissociation-bond formation for hydrogen are given by,
\begin{equation}
  \begin{aligned}
  & H + e^{-} \rightleftharpoons H^{+} + 2e^{-} \\
  & H_{2} \rightleftharpoons H + H,
  \end{aligned}
\end{equation}
respectively.
Following \cite{DAngelo2013}, we define the degree of dissociation $y$ and degree of
ionization $x$ as,
\begin{equation}\label{eq:8}
\begin{array}{l}
  \DS y =\frac{\rho_{HI}}{\rho_{HI} + \rho_{H_2}} \\ \noalign{\medskip}
  \DS x =\frac{\rho_{HII}}{\rho_{HI} + \rho_{HII}},
\end{array}
\end{equation}
where, $\rho_{HI}$ is the density of atomic hydrogen,  $\rho_{H_2}$ the density of molecular hydrogen and $\rho_{HII}$ the density of ionized hydrogen.
In the limit of LTE, one assumes that the level populations due to ionization (and dissociation) processes follow Boltzmann excitation formula and that the ejected free electrons thermalize to attain a Maxwell-Boltzmann velocity distribution corresponding to single gas temperature.
This is generally true in regions of high density like that of the solar interior. In such cases, the degree of ionization using Saha equations is given as follows,
\begin{equation}\label{eq:9}
\frac{x^2}{1 - x} = \frac{m_H}{X\rho}\left(
  \frac{m_ek_BT}{2\pi\hbar^2}\right )^{3/2} e^{-13.60 eV/(k_B T)}\,,
\end{equation}
and also degree of dissociation, $y$ can be obtained in a similar manner \citep{Black1975},
\begin{equation}
\label{eq:10}
\frac{y^2}{1 - y} = \frac{m_H}{2X\rho}\left(
  \frac{m_Hk_BT}{4\pi\hbar^2}\right )^{3/2} e^{-4.48 eV/(k_B T)}\,.
\end{equation}

The gas is essentially a mixture of hydrogen in all forms (atoms, ions \& molecules) with a mass fraction of $X$, Helium with a mass fraction of $Y$ and negligible fraction of metals.
For such a composition the total density of gas is defined as $\rho = n\mu m_H$, where the mean molecular weight $\mu$ can be expressed as \citep[e.g.,][]{Black1975}  
\begin{equation}
\label{eq:11}
\frac{\mu}{4}= \left[2\textit{X} (1 + y + 2xy) + \textit{Y}\right]^{-1}.
\end{equation}
Such a gas mixture is further assumed to be thermally ideal so that pressure and temperature are related by $p = \rho k_B T/(\mu m_H)$.

The most crucial part is to express a caloric \textit{EoS} that can account for contributions from various degrees of freedom and processes like ionization and dissociation.
Thus, the gas internal energy density $(\rho e)_{gas}$ for the mixture is given by
\begin{equation}\label{eq:H-HeEoS}
(\rho e)_{gas} = (\epsilon_{H_2} + \epsilon_{HI} + \epsilon_{HII}
                   + \epsilon_{H+H} + \epsilon_{He})
                 \frac{\rho k_B T}{m_H}\,, 
\end{equation}
where each term in parenthesis is dimensionless and can be obtained from an appropriate partition function $\mathcal{Z}$ and Eq. (\ref{eq:ideal}).
Table \ref{tab:H-HeEoS} summarizes the different contribution to the gas internal energy.

\begin{table*}
\begin{center}
\caption{\footnotesize Summary of different contributions to the gas internal energy
  $(\rho e)_{gas}$, which is expressed using Eq. (\ref{eq:H-HeEoS}. (see
  \cite{Black1975, DAngelo2013})}
\label{tab:H-HeEoS}
\begin{tabular}{c c c}
\hline
Term & Expression & Description \\
\hline
\smallskip
$\epsilon_{HI}$ & 1.5X\,(1 + x)\,y & Translational energy for hydrogen \\
\smallskip
$\epsilon_{He}$ & 0.375Y & Translational energy for helium \\
\smallskip
$\epsilon_{H+H}$ & $4.48\,{\rm{eV}}\,X\,y /(2 k_B T)$ & Dissociation
energy for molecular hydrogen \\
\smallskip
$\epsilon_{HII}$ & $13.6\,{\rm{eV}}\,X\,x\,y / (k_B T)$ & Ionization 
energy for atomic hydrogen \\
\smallskip
$\epsilon_{H2}$ & $\frac{X\,(1-y)}{2}\left[1.5 + \frac{T}{\zeta_v} \frac{d
    \zeta_v }{dT} + \frac{T}{\zeta_r} \frac{d \zeta_r}{dT}\right] $& Internal energy for
molecular hydrogen \\
\hline
\end{tabular}
\end{center}
\end{table*}

In the case of molecular hydrogen, $\epsilon_{H_2}$, terms that correspond to vibrational and rotational degree of freedom are also considered.
These terms are evaluated using the partition function of vibration $\zeta_v$ and rotation $\zeta_r$ that have explicit and a non-linear dependence on temperature.
Additionally, the rotational partition function also takes into account the para/ortho H$_2$ spin states \citep{Boley2007}.
Thus, the total gas internal energy density has a nonlinear dependence on the temperature T and density through $x$ and $y$ (see Eqns. \ref{eq:9} \& \ref{eq:10}). 

The left and middle panels of Fig. \ref{fig:muintegmm} show the variation of $\mu (\rho, T)$ (Eq. \ref{eq:11})  and gas internal energy in ergs with temperature, T, for four values of density in g\,cm$^{-3}$ respectively.
The values of $\mu$ are bounded between the upper value $\sim$2.3, corresponding to a fully molecular medium at low temperatures and a lower value $\sim$ 0.6 at high temperatures representing a fully ionized medium.
The transition between these bounds is smooth at large densities $\rho = 10^{-4}$ g\,cm$^{-3}$ while it forms an intermediate plateau at $T \sim 10^{3}$ K at low density values  (black curve).
The first transition occurs in the temperature range where molecules begin to dissociate to form atomic hydrogen.
A second transition takes place where atomic hydrogen becomes ionized.
The same transitions can be observed in the profile of internal energy.
From a physical point of view they indicate that the energy at these temperatures becomes available to dissociate or ionize the gas rather than heating the gas so that temperature remains approximately constant. 
Away from these transition regions, the dependence of $(\rho e)_{\rm gas}(m_H/\rho)$ is linear and increases monotonically with the gas temperature.
The last panel of the same figure shows the variation of first adiabatic exponent, $\Gamma_1$ with temperature.
At low temperatures, the gas behaves as a monoatomic ideal gas undergoing adiabatic process with $\Gamma_1 = 5/3$.
This is also true at very large temperatures where the gas contains ions and electrons.
From the previous considerations, we see a sharp decrease in $\Gamma_1$ from its maximum value of $5/3$ to values around unity (corresponding to an isothermal limit) for a low density plasma (black curve).
On the other hand, a single dip at $T > 10^{4}$ K is seen at larger densities (red curve).

 In addition to the study of planet formation in accretion
  discs, the H-He EoS is an
  important ingredient in the physics of proto-stellar formation from
  collapse of dense molecular cores. Radiation hydrodynamics
  simulations \citep{Masunaga1998, Masunaga2000} have put forth detailed understanding of
  thermodynamics in presence of gravitational collapse. At the onset of
  collapse, compressional heating in the dense ($\rho \sim$ 10$^{4-5}$ g
  cm$^{-3}$)  and cold (T $\sim$ 10\,K) core increases the central
  temperature up to T $\sim$ 100\,K adiabatically. As the
  temperature increases further, rotational states of H$_{2}$ are
  excited and the system evolves with an effective adiabatic
  index of 7/5 with the ensuing formation of a pressure-supported 
  \textit{first core}. Further increase of temperature beyond 10$^{3}$\,K results in dissociation
  of H$_{2}$ which acts as an efficient cooling mechanism leading to a
  second collapse.

  However, not all astrophysical problems can be treated in LTE limit.
A classical case is that of a jet, where the recombination time scales are comparable to that of dynamical time.
In such a scenario, LTE assumptions become invalid and a non-equilibrium approach has to be adopted as described in the following section. 

\subsection{Non-Equilibrium Hydrogen Chemistry}
\label{sec:h2cool}
%
%
Astrophysical flows in HII regions, supernova remnants, star forming regions are some classical examples where optically thin cooling time scales are comparable to the dynamical time. 
In such environments, ionization and dissociation fractions are far from LTE and their estimation based on Saha fractions can give large errors.
In such cases, the number density of various species is more accurately determined by solving the chemical rate equations:
\begin{equation}\label{eq:chemical_reaction}
\frac{dn_{i}}{dt} = \sum_{j,k} \mathcal{K}_{j,k} n_{j} n_{k} -
                    n_{i}\sum_{j}\mathcal{K}_{i,j} n_{j},
\end{equation}
where $n$ is the number density, $\mathcal{K}_{j,k}$ is the rate of formation of $i^{th}$ specie from all $j$ and $k$ species while $\mathcal{K}_{i,j}$ is the rate of destruction of the $i^{\rm th}$ specie due to all $j$ species.

In addition, proper treatment should be carried out to evolve the internal energy to account for losses due to optically thin radiation:
\begin{equation}\label{eq:cooling}
  \frac{d(\rho e)}{dt} = - \Lambda(n, \vec{X}, T)\,,
\end{equation}
where $\Lambda(n, \vec{X}, T)$ is the optically thin radiative loss term.
Radiative losses imply that the emitted photons due to different physical processes (e.g., ionization, metal line cooling etc.) freely stream (without diffusion) away from the region where they are produced and eventually escape into the surroundings resulting into an effective
decrease in total gas internal energy.

In presence of cooling, the gas internal energy, $(\rho e)_{\rm gas}$, will be different from that defined by Eq. (\ref{eq:H-HeEoS}).
Indeed, only contributions due to translational and internal degrees of freedom (from $H$, $He$ and $H_{2}$) should be included.
Conversely, terms corresponding to the emission of photons (e.g. ionization, dissociation, roto-vibrational cooling of $H_{2}$ molecule) are correctly accounted for by the right hand side of Eq. (\ref{eq:cooling}) in the $\Lambda$ term.
Therefore Eq. (\ref{eq:H-HeEoS}) now becomes
\begin{equation}\label{eq:HHe_cool}
(\rho e)_{gas} = (\epsilon_{H2} + \epsilon_{HI} + \epsilon_{He})\frac{\rho k_B T}{m_H},
\end{equation}
where, expressions for each of the internal energy components are given in table \ref{tab:H-HeEoS}.
A similar contribution to gas internal energy (see. Eq. (\ref{eq:cooling})) from internal degrees of freedom in
presence of radiative losses has also been applied to study the role of molecular hydrogen in primordial star formation \citep[e.g.,][]{Palla:1983aa, Omukai1998}. 

For the present purpose, we focus only in the chemical evolution of atomic and molecular hydrogen. 
In particular, the total hydrogen number density $n_{\rm H}$ includes contributions from atomic and molecular hydrogen i.e., $n_H$ = $n_{HI} + 2n_{H_2} + n_{HII}$. 
Contributions to the electrons density $n_{e}$ come from ionized hydrogen ($n_{HII}$) and from a small but fixed fraction of metals ($Z \sim 10^{-4}$). 
In addition to hydrogen, we also consider helium to be present with a fixed mass fraction of $0.027$. 
The mass density $\rho
  = \mu n_{\rm tot}\, m_{p}$ is a conserved quantity 
  whereas the total number of particle per unit volume ($n_{\rm tot} =
  n_H + n_{He} + n_{e}$) is not as it may change due to ionization, recombination
  and dissociation processes.
The chemical evolution of molecular, atomic and ionized hydrogen is governed by the equations listed in Table \ref{tab:chemeq}.  
The code tracks the formation and destruction of these three species based on the temperature-dependent reaction rates specified and updates the corresponding number fractions,
\begin{equation}
  X_{HI}  = \frac{n_{HI}} {n_H}\,;\qquad
  X_{H_2} = \frac{n_{H_2}}{n_H}\,;\qquad
  X_{HII} = \frac{n_{HII}}{n_H}\,.
\end{equation}

\begin{table*}
\begin{center}
\caption{\footnotesize Summary of the chemistry reaction set. T is the temperature
  in Kelvin, $T_{\rm eV}$ is the temperature in electron-volts, $T_{5}$
  = $T/1\times10^{5}$  and $T_{2}$  = T/100}
\label{tab:chemeq}
\begin{tabular}{l l l l}
\hline
No. & Reaction & Rate Coefficient ($\rm {cm}^{3} s^{-1}$) &
Reference\\
\hline
1. & H + e$^{-}$ $\rightarrow$ H$^{+}$ + 2e$^{-}$ & $k_1$ = $5.85
\times 10^{-11}$ $T^{0.5}$ \rm{exp}(-157,809.1/T)/(1.0 +
$T_{5}^{0.5}$) & 1\\
2. & H$^{+}$ + e$^{-}$ $\rightarrow$ H + h$\nu$ & $k_2$ =
$3.5\times10^{-12} (T/300.0)^{-0.8}$ & 2\\
3. & H$_{2}$ + e$^{-}$ $\rightarrow$ 2H + e$^{-}$ & $k_3$ =
$4.4\times10^{-10} T^{0.35} \rm{exp}(-102,000.0/T)$ & 3\\
4. & H$_{2}$ + H $\rightarrow$ 3H & $k_4$ = $1.067\times10^{-10}
T_{\rm eV}^{2.012}(\rm{exp}(4.463/T_{\rm eV})^{-1}((1. + 0.2472 T_{\rm eV})^{3.512})^{-1} $& 4\\
5. &H$_{2}$ + H$_{2}$ $\rightarrow$ H$_{2}$ + 2H & $k_5$ = $1.0\times
10^{-8} \rm{exp}(-84,100/T)$ & 2\\
6. & H + H $\overset{\rm dust}\longrightarrow$ H$_{2}$ & $k_6 =
3.0\times10^{-17}\sqrt{T_{2}}(1.0 + 0.4\sqrt{T_{2} + 0.15} + 0.2T_{2} + 0.8T_{2}^{2})$ & 5 \\
\hline

\end{tabular}
\tablebib{(1) \cite{Cen1992} [Eq. 26a];
  (2) \cite{Woodall2007} [UMIST Database] (3)
  \cite{Galli1998} [Eq. H17]; (4) \cite{Abel1997}
  [Tab. 3 Eq. 13]; (5) \cite{Hollenbach1979} [Eq. 3.8]}
\end{center}
\end{table*}

\begin{figure}
\centering
\includegraphics[width=1.0\columnwidth]{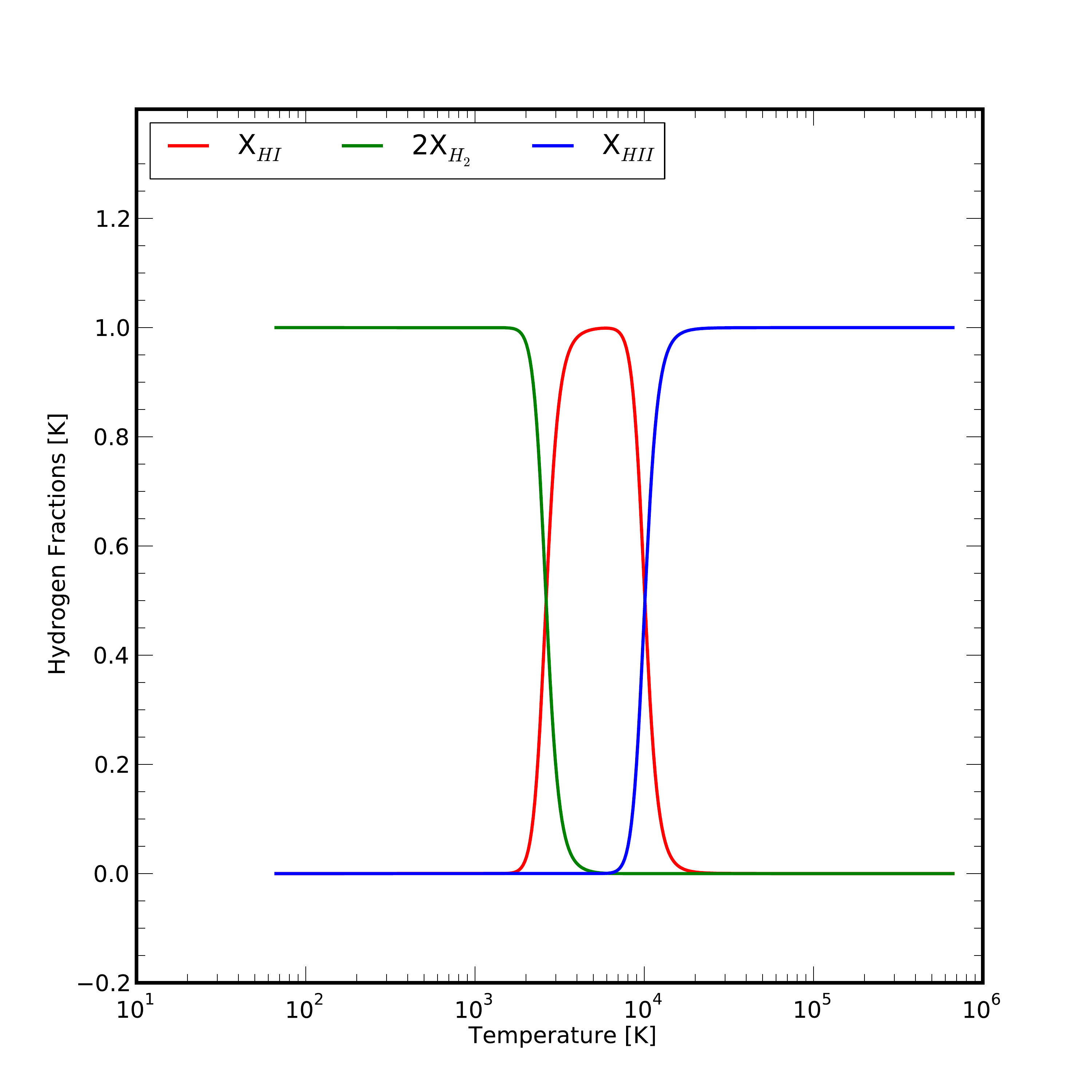}
\caption{\footnotesize Different hydrogen fractions ($X_{HI}$ : Atomic \textit{red},
  $X_{H_2}$ : Molecular \textit{green} and $X_{HII}$ : Ionized \textit{blue})  obtained at
  \textit{equilibrium} for different temperatures. 
  Note that the total sum of fractions, i.e, $X_{HI} + 2X_{H_2} + X_{HII}$ is 
  conserved.}
\label{fig:h2cooleq}
\end{figure}

In dilute regions such as the solar corona, Eq. (\ref{eq:chemical_reaction}) can be simplified by setting $dn_{i}/dt=0$ as the time scales are such that a balance is always maintained between collisional ionization and radiative recombination.
This condition is known as {\em Coronal equilibrium} or {\em Collisional Ionization Equilibrium} (CIE) and it differs from LTE in two aspects: i) it is only valid in dilute plasma - unlike the LTE where high density environments are required - and ii) the ionization fraction are estimated using Eq. (\ref{eq:chemical_reaction}) in steady state and not with Saha fractions.
The top panel of Fig. \ref{fig:h2cooleq} shows the concentration fractions as functions of temperature obtained by solving the steady state version of Eq. (\ref{eq:chemical_reaction}).
The dissociation and ionization of molecular and atomic hydrogen, respectively, are clearly evident at the temperatures of $T\sim 3\times 10^{3}$\,K and $T\sim 10^{4}$\,K.
Such equilibrium values can be used to initialize fractions in numerical computations.

The radiative losses implemented in our model can be written as
\begin{equation} \label{eq:17}
\Lambda(n,\vec{X},T) =   \Lambda_{\rm CI} + \Lambda_{\rm RR}
                       + \Lambda_{\rm rotvib} + \Lambda_{\rm H2diss} 
                       + \Lambda_{\rm grain}\,,
\end{equation}
where, $\Lambda_{\rm CI}$  and $\Lambda_{\rm RR}$ are losses due to collisional ionization and radiative recombination, respectively \citep{Tesileanu2008}.
The remaining terms, $\Lambda_{\rm rotvib}$, $\Lambda_{\rm H2diss}$ and $\Lambda_{\rm grain}$ are associated with molecular hydrogen and represent losses due to rotational-vibrational cooling, dissociation and gas-grain processes \citep{Smith2003}.
\begin{figure}
\centering
\includegraphics[width=1\columnwidth]{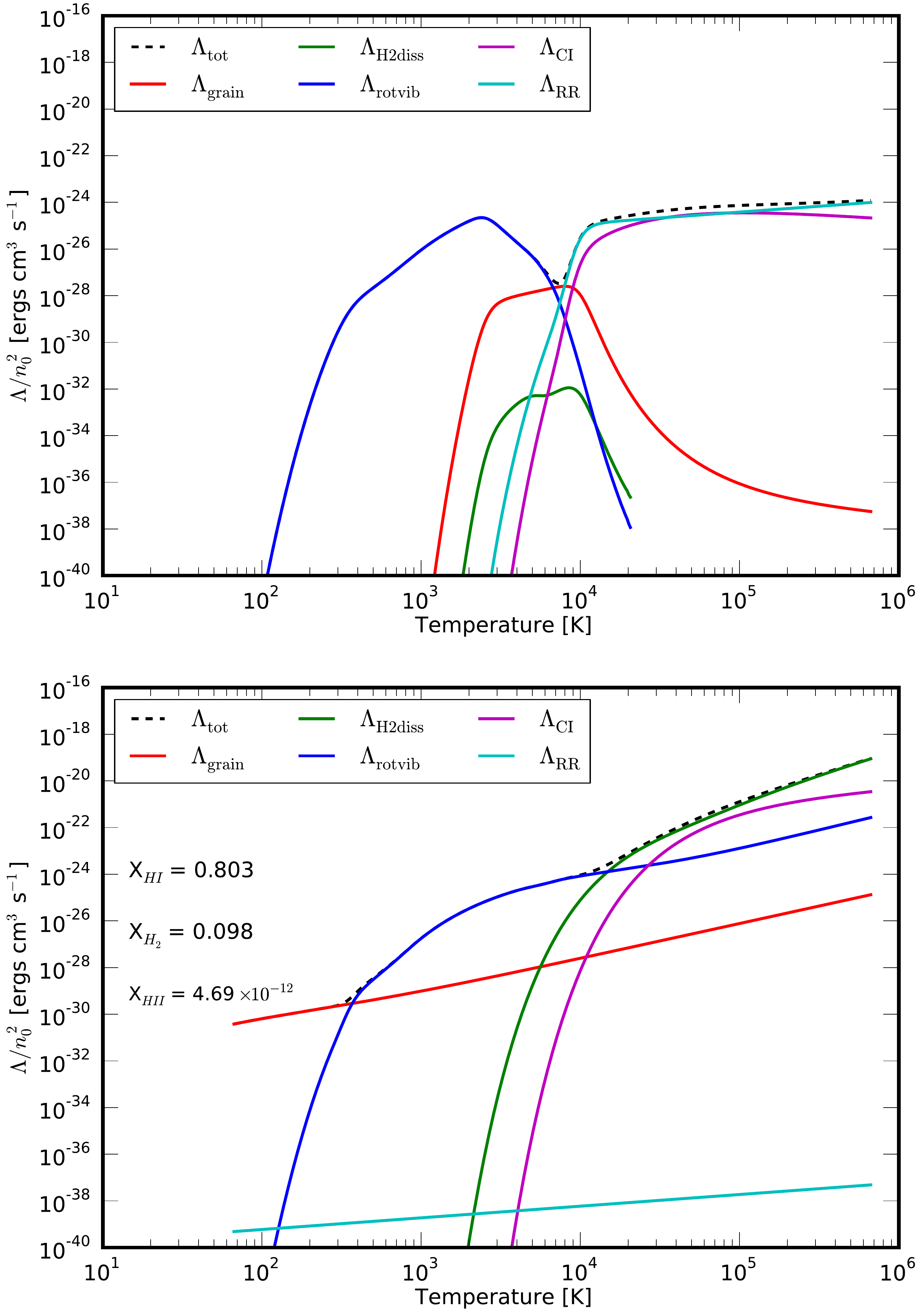}
\caption{\footnotesize 
  \textit{Top Panel:} Variation with temperature of cooling function
  arising from different processes obtained using
  equilibrium values of hydrogen fractions (that vary with temperature
  as shown in Fig. \ref{fig:h2cooleq}) in 
  units of ergs cm$^{3}$ s$^{-1}$ for a total number
  density $n_0 = 10^{5}$ cm$^{-3}$. 
 \textit{Bottom panel:} Different components of radiative cooling
  functions with same initial number density, n$_0$ 
  but for fixed values of concentration fractions (mentioned in the
  figure). In both panels, the sum of all the components
  is drawn with a {\em black dashed} line to obtain the value of $\Lambda$ in
  ergs\,cm$^{3}$\,s$^{-1}$ following Eq. (\ref{eq:17}).}
\label{fig:lambdafunc}
\end{figure}

The dependence of the cooling function, $\Lambda(n,\,\vec{X},\, T)/n_0^{2}$, for $n_0 = 10^{5}$ cm$^{-3}$ is shown in the top and  bottom panels of Fig. \ref{fig:lambdafunc}.
In the top panel, cooling functions are plotted using concentrations values obtained under  CIE conditions while the plots in the bottom panel corresponds to fixed concentrations obtained for $T = 4500$ K.
For temperatures $T < 10^{4}$ K, the total cooling function with equilibrium values of hydrogen fractions (\textit{black dashed curve}) is dominated by the contribution of roto-vibrational cooling of molecular hydrogen (\textit{blue line}). 
Molecular cooling due to dissociation (\textit{green curve}) and gas-grain processes (\textit{red curve}) have little impact on the total cooling function.
For temperatures  $T> 10^{4}$ K, most of the hydrogen molecules have
dissociated into atoms (see Fig.\ref{fig:h2cooleq}) and the total
cooling curve in Fig. \ref{fig:lambdafunc} is governed by atomic
processes like radiative recombination (\textit{cyan line})  and collisional ionization (\textit{magenta line}) . 

The total cooling curve in the bottom panel of Fig. \ref{fig:lambdafunc} with fixed values of hydrogen fractions
differs substantially from the panel above. It is dominated by gas grain processes (\textit{red line}) for $T< 100$ K.
For temperatures between $100$ K and $10^{4}$ K, molecular cooling due to rotational and vibrational processes (\textit{blue line}) plays a vital role.
Even at large temperatures, $T> 10^{4}$ K,  the total cooling curve has major contribution from molecular dissociation (\textit{green line}), for the chosen fixed fraction of molecules. 
The contribution of collisional ionization (\textit{magenta line}) also becomes important at these temperatures.
However, the cooling due to radiative recombination remain negligible for all temperature values due to extremely small and fixed fraction of electrons, $X_{HII}$.

\section{Numerical Implementation}
\label{sec:numframe}
%
%
%

In a Godunov-type scheme the MHD equations (\ref{eq:mhd_rho})--(\ref{eq:mhd_X}) are discretized using a flux-conserving form where the basic building block is
\begin{equation}\label{eq:conservative_update}
  \vec{U}^{n+1}_{\vec{i}} = \vec{U}^n_{\vec{i}} - \frac{\Delta t^n}{\Delta{\cal V}_{\vec{i}}}
                  \sum_d\Big[\left(A_d\vec{F}_d\right)_{\vec{i}+\HALF\hvec{e}_d}
                            -\left(A_d\vec{F}_d\right)_{\vec{i}-\HALF\hvec{e}_d}\Big]
                  + \vec{S}_{\vec{i}}
\end{equation}
where $\vec{U} = (\rho, \rho \vec{v},\, \vec{B},\, E,\, \rho X_k)$ is our vector of conservative variables, $\Delta t^n$ is the time step and $\vec{S}_{\vec{i}}$ is a source term.
Here we employ an orthogonal system of coordinates with unit vectors $\hvec{e}_d$ (here $d=1,2,3$ or simply $x,y,z$) where $A_d$ is the area element in the $d$ direction, $\vec{F}_d$ is the flux computed with a Riemann solver, ${\cal V}_{\vec{i}}$ is the cell volume while $\vec{i}=(i,j,k)$ is the position of the computational zone in the domain.
We remind the reader to \cite{Mignone2007}, \cite{Tesileanu2008} and \cite{Mignone2012} for more details.
Here we focus only on those aspects that crucially depend on the choice of the equation of state, namely: i) the computation of the flux via a Riemann solver and ii) the conversion between internal energy and pressure.

\subsection{On the solution of the Riemann Problem}
%
%
%

The Riemann problem is an initial value problem describing the decay of a discontinuity separating two constant states.
As time evolves, the discontinuity breaks into a set of non-interacting elementary waves whose properties are determined by connecting, in a self-consistent way, the initial left and right states through wave-curves.
For a convex EoS, as defined in Sect. \ref{sec:thermodynamic_constraints}, the solution to the Riemann problem in hydrodynamics consists of a left-facing shock or rarefaction wave, a contact discontinuity and a right-facing wave (again either a shock or a rarefaction).
No compound wave can be formed in the solution.
In MHD the solution may involve up to 7 modes:  a pair of fast magneto-sonic waves, a pair of Alfv$\acute{e}$n waves (or rotational modes) and a pair of slow waves and a contact (or tangential) discontinuity in the middle.
Although exact solutions to the Riemann problem are possible, the computational overhead is largely reduced by using approximate solvers based on different levels of simplification.

A first class of solvers heavily relies on characteristic (Jacobian) decomposition or computation of Riemann invariants which is strictly connected with the underlying form of the conservation law.
Typical examples are linearized (Roe-type), flux-splitting or two-shock approximate solvers, see the book by \cite{Toro2009}.
Solvers belonging to this class require considerable changes when new physical ingredients  (such as a different EoS) are introduced.
In the case of real gases, for instance, extensions have been presented by \cite{Colella1985, Glaister1988a, Glaister1988b, Buffard2000} (in the case of the Euler or Navier-Stokes equations) while generalization to the MHD case have been presented in \cite{Dedner2001}.
In \cite{Fedkiw1997} a Roe-type Riemann solver for the solution of thermally ideal, chemically reacting gases has been presented in the context of the multi-species Navier-Stokes equations.

A second class of solvers employs only minimal information (typically approximate expressions for the wave speeds) and it is based on an application of the integral form of the conservation laws which gives a closed-form approximate expression for the fluxes.
Typical examples are the Lax-Friedrichs-Rusanov \citep{TothOdstricil.1996}, Harten-Lax-van Leer \citep[HLL, see][]{HLL83} solver and their extensions such as HLLC \citep{Toro1994, Gurski.2004,Li2005}  and HLLD \citep{Miyoshi2005}.  
Solvers belonging to this class can accomodate new changes with minimal efforts by simply changing the definition of the sound speed and, for this reason, will be our preferred method of choice.
Incidentally, we also note that, since only an approximate expression for the eigenvalue is needed and $\Gamma_1 \le 5/3$, employing Eq. (\ref{eq:cs}) with $\Gamma_1 = 5/3$ provides an upper bound to the actual expression with a only a slight loss of accuracy.

\subsection{Conversion between internal energy and pressure}
%
%
%

\begin{table*}[!ht]
\caption{\footnotesize Equations being solved when converting from pressure to
internal energy ($p\to\rho e$) or viceversa ($\rho e\to p$).
Equations marked with a * are nonlinear and may require a root-finder approach.
Under non-equilibrium conditions the chemical fractions $\vec{X}$ are independent variables available immediately after the hydro update.
Under LTE or CIE, $\vec{X}=\vec{X}(T,\rho)$ and this introduces additional nonlinearities in some equations.}
\label{tab:conversion}
\centering
\begin{tabular*}{0.95\textwidth}{l@{\extracolsep{\fill}}lll}\hline
  Conversion      &  Physical Conditions  &     Temperature     &    Final Equation\\ \hline\hline \noalign{\medskip}
 $(\rho, \vec{X}, p)\to\rho e$         &  [Non-equilibrium] &
 $\DS T = \frac{p}{\rho}K\mu(\vec{X})$ &
 $\DS   \rho e = \rho e(T,\vec{X})$
   \\ \noalign{\smallskip}\hline\noalign{\smallskip}
 $(\rho, \vec{X}, \rho e)\to p$         & [Non-equilibrium] &
 $\DS \rho e(T,\vec{X}) - \rho e = 0^*$ &
 $\DS p = \frac{\rho T}{K\mu(\vec{X})}$
   \\ \noalign{\smallskip}\hline\noalign{\smallskip}
 $(\rho,  p)\to\rho e$                       & [LTE or CIE] &
 $\DS T - \frac{p}{\rho}K\mu(T,\rho) = 0^*$ &
 $\DS   \rho e = \rho e(T,\rho)$       
   \\ \noalign{\smallskip}\hline\noalign{\smallskip}
 $(\rho, \rho e)\to p$                  &  [LTE or CIE] &
 $\DS \rho e(T,\rho) - \rho e = 0^*$    & 
 $\DS  p = \frac{\rho T}{K\mu(T,\rho)}$
   \\ \noalign{\smallskip}\hline\noalign{\smallskip}
\end{tabular*}
\end{table*}

The computation of the right hand side of Eq. (\ref{eq:conservative_update}) is normally carried out using primitive variables customarily defined as $\vec{V}=(\rho, \vec{v}, \vec{B}, p)$.
The conversion between $\vec{U}$ and $\vec{V}$ requires obtaining pressure from total energy density and viceversa.
While internal energy density is readily obtained from Eq. (\ref{eq:total_energy}), the conversion $p\to \rho e$ and its inverse $\rho e\to p$ strictly depends on the choice of the caloric equation of state.

For the constant-$\Gamma$ EoS, these transformations take a small fraction of the computational time as the relation between internal energy and pressure is straightforward and given by
\begin{equation}
 \rho e = \frac{p}{\Gamma - 1}\,.
\end{equation}
Note also that the temperature does not explicitly appears in the previous definition.

The situation is different, however, for a more general EoS where a closed-form expression between pressure and internal energy may not be easy to obtain.
From the considerations given in the previous sections, in fact, we can write the thermal and caloric equations of state as
\begin{equation}\label{eq:thermal+caloric}
  \left\{\begin{array}{lcl}
     p & = & \DS\frac{\rho k_BT}{m_u \mu(\vec{X})} \\ \noalign{\medskip}
     e & = & e(T,\vec{X})
  \end{array}\right.
\end{equation}
where $\vec{X}$ may be an independent variable or, in equilibrium conditions, a function of temperature and density.
The explicit dependence on the temperature introduces two additional intermediate steps, namely:
\begin{enumerate}
  \item
   During the conversion $p\to\rho e$ one first needs to compute $T$ from
   the thermal EoS (first of Eq. \ref{eq:thermal+caloric}):
   \begin{equation}\label{eq:PV_Temp}
     T = \frac{p}{\rho} \frac{m_u\mu(\vec{X})}{k_B} \,.
   \end{equation}
   Under non-equilibrium conditions, $\mu=\mu(\vec{X})$ is a known function of
   the gas composition and Eq. (\ref{eq:PV_Temp}) can be solved directly.
   Under LTE or CIE, on the other hand, $\vec{X} = \vec{X}(T,\rho)$ is
   a function of density and temperature and Eq. (\ref{eq:PV_Temp})
   becomes a nonlinear equation in the temperature variable.
 
  \item
   During the inverse transformation ($\rho e\to p$), one must first solve for
   the temperature by inverting
   \begin{equation}\label{eq:EV_Temp}
     e = e(T,\vec{X}) \,,
   \end{equation}
   where, under equilibrium assumptions, the chemical composition is a function
   of temperature and density, i.e., $\vec{X}=\vec{X}(T,\rho)$.
\end{enumerate}
Table \ref{tab:conversion} summarizes the relevant equations to be solved.

In order to cope with the numerical inversion of Eqns. (\ref{eq:PV_Temp}) and (\ref{eq:EV_Temp}) we have considered and implemented two different solution strategies that we describe in the following sections.

\subsubsection{Conversion using root finders}
\label{sec:root_conversion}
%

In this approach we employ the exact analytical expressions (\ref{eq:PV_Temp})-(\ref{eq:EV_Temp}) to compute pressure and internal energy as functions of temperature or viceversa.
For the caloric EoS, numerical inversion using a root-finder algorithm is required to obtain $T$ from $\rho e(T,\vec{X})$ or $\rho e (T,\rho)$ as they both are nonlinear functions of the temperature.
Additionally, under LTE or CIE, the thermal EoS must also be inverted numerically to obtain temperature since the mean molecular weight introduces nonlinearity.
These cases are marked with a * in Table \ref{tab:conversion}.

The root-finder approach results in increased computational cost inasmuch the internal energy is an expensive function to evaluate. 
Among different root-solvers not requiring the knowledge of the derivative, we have found Brent's or Ridders' methods to be practical and efficient root-finding algorithms.

\subsubsection{Conversion using Tables}
\label{sec:table_conversion}
%

A second and more efficient strategy can be used when the internal energy is a function of temperature and density alone (which is typically the case under equilibrium conditions, CIE or LTE) and it consists of employing pre-computed tables of pressure and internal energy, e.g.,
\begin{equation}
 \{p\}_{ij} = p(T_i, \rho_j) \,,\qquad
 \{\rho e\}_{ij} = \rho e(T_i, \rho_j)
\end{equation}
where $i=0,...,N_c-1$ and $j=0,...,N_r-1$ are the table indices.
For convenience, the tables are constructed using equally-spaced node values in $\log T$ and $\log\rho$ so that $\log T_i/T_0 = i\Delta\log T$ and $\log \rho_j/\rho_0 = j\Delta\log\rho$ where $\rho_0$ and $T_0$ are the lowest density and temperature values in the table.
Following this approach, Eqns. (\ref{eq:PV_Temp})-(\ref{eq:EV_Temp}) and their inverses are replaced by direct or reverse lookup table operations.
We point out that, in order to be invertible, tables must be obtained from monotone functions of their arguments and this is always verified for the EoS of interest.

When $T$ and $\rho$ are known, for instance, pressure and internal energy are obtained by using lookup table followed by two-dimensional interpolation between adjacent node values.
Thus we first locate the table indices $i$ and $j$ by a simple division:
\begin{equation}\label{eq:tab_indices}
  i = \mathrm{floor}\left(\frac{\log T/T_0}{\Delta \log T}\right)
  \,,\quad
  j = \mathrm{floor}\left(\frac{\log\rho/\rho_0}{\Delta\log \rho}\right)
\end{equation}
where $T$ and $\rho$ are the input values at which interpolation is desired.
Internal energy (and similarly pressure) is then computed as
\begin{equation}\label{eq:lookup_table}
    \rho e(T,\rho) = {\cal S}_{i,j}(x) (1 - y) + {\cal S}_{i,j+1}(x) y
\end{equation}
where $x = (T-T_i)/(T_{i+1}-T_i)$ and $y = (\rho - \rho_j)/(\rho_{j+1}-\rho_j)$ are normalized coordinates between adjacent nodes while ${\cal S}$ is an interpolating spline, see Sect. \ref{sec:interpolation}.

Conversely, when $\rho e$ and $\rho$ are known, Eq. (\ref{eq:lookup_table}) must be inverted to obtain $T$. 
To this end, we first locate the index $j$ using the second of (\ref{eq:tab_indices}) with $\rho$ given.
In order to find the column index $i$, we first construct the intermediate one-dimensional array $q_i = \{\rho e\}_{i,j}(1-y) + \{\rho e\}_{i,j+1}y$ with $j$ fixed from the previous search whereas $i=0,...,N_c$.
A binary search is then performed on $q_i$ to obtain $i$ and Eq. (\ref{eq:lookup_table}) can be inverted to obtain $x$.
Temperature is finally obtained as $T = T_i + x(T_{i+1}-T_i)$.
The inversion depends on the choice of the interpolating function ${\cal S}$ and it is discussed in more detail in the next section.

The tabulated approach has found to be faster than the general root finder method giving considerable speedups, up to a factor of 4 for certain problems. 

\subsubsection{On the choice of the interpolant}
\label{sec:interpolation}
%

For linear interpolation, we simply use
\begin{equation}\label{eq:linear_spline}
 {\cal S}_{i,j}(x) = \{\rho e\}_{i+1,j} x_i + \{\rho e\}_{i,j} (1-x_i) 
\end{equation}
in Eq. (\ref{eq:lookup_table}) so that internal energy and pressure become bilinear functions of temperature and density.
When $\rho e$ is given, on the other hand, the inverse function is required to compute temperature and one needs to solve Eq. (\ref{eq:lookup_table}) with ${\cal S}(x)$ given by Eq. (\ref{eq:linear_spline}) to obtain $x$.
Since the resulting equation is linear in $x$ one readily finds:
\begin{equation}\label{eq:linear_sol}
   x = \frac{f - f_{i,j} - y(f_{i,j+1}-f_{i,j})}
            {y(f_{i+1,j+1} - f_{i,j+1} - f_{i+1,j} + f_{i,j}) + f_{i+1,j}-f_{i,j}}
\end{equation}
where $f = \rho e$, $f_{i,j} = \{\rho e\}_{i,j}$ or $f = p$, $f_{i,j} = \{p\}_{i,j}$.
Despite its simplicity, however, bilinear interpolation may generate thermodynamically inconsistent results since the positivity of the fundamental derivative (Eq. \ref{eq:G}) can be violated at node points in the table where derivatives are not continuous.
This generates composite waves in the solution due to a loss of convexity of the caloric EoS as discussed in Sect. \ref{sec:thermodynamic_constraints}.
Indeed, we have verified that this pathology worsen as the number of points in the temperature grid is reduced.
An illustration of the problem is reported in Appendix \ref{sec:Gtest}.

To overcome this limitation, we have also implemented a cubic spline when interpolating along the temperature grid so that
\begin{equation}\label{eq:cubic_spline}
  {\cal S}_{i,j}(x) = a_{i,j}x^3 + b_{i,j}x^2 + c_{i,j}x + d_{i,j}\,,
\end{equation}
where the coefficients $a,b,c$ and $d$ are computed by ensuring that the cubic is strictly monotonic in the interval $[T_i,T_{i+1}]$ see Appendix \ref{sec:monotone_spline}.
The inverse function requires finding the root of a cubic equation on a specific interval and, since the spline is monotone by construction, only one root is always guaranteed to exist.
Thus combining Eq. (\ref{eq:cubic_spline}) with Eq. (\ref{eq:lookup_table}) one obtains the following equation:
\begin{equation}
  h(x) =   \tilde{a}_{i,j} x^3
         + \tilde{b}_{i,j} x^2
         + \tilde{c}_{i,j} x
         + \tilde{d}_{i,j} - f = 0 
\end{equation}
where $\tilde{a}_{i,j} = a_{i,j}(1-y) + a_{i,j+1}y$ and similarly for the remaining coefficients.
Although standard analytical solvers (e.g. Cardano) may be used, we have found root-finder methods to be more robust, accurate and computational efficient for the purpose.
Starting from an initial guess given by Eq. (\ref{eq:linear_sol}), one can either use Newton-Raphson method to achieve quadratic convergence,   
\begin{equation}
  x^{[k+1]} = x^{[k]} - \frac{h(x^{[k]})}{h'(x^{[k]})}  \,,
\end{equation}
or Halley's method which gives cubic convergence:
\begin{equation}
  x^{[k+1]} = x^{[k]}
  - \frac{2h(x^{[k]})h'(x^{[k]})}{2[h'(x^{[k]})]^2 - h(x^{[k]})h''(x^{[k]})} \,.
\end{equation}
We have found that both methods hardly never require more than $3$ iterations to achieve an absolute accuracy of $10^{-11}$.

\section{Numerical benchmarks}
\label{sec:tests}
%
%
%

The computational framework presented in this work has been implemented in the PLUTO code \citep{Mignone2007, Mignone2012}.
Selected numerical benchmarks are now presented with the aim of investigating the influence of the EoS on problems of astrophysical relevance as well as comparing the computational efficiency of the proposed numerical approaches.

\subsection{Sod shock tube}
\label{sec:sodtest}
%
%

In our first test, we consider the standard Sod shock tube on the unit interval $x\in[0,1]$ with a uniform grid resolution $\Delta x = 10^{-3}$. 
The initial setup consists of two fluids initially separated by a discontinuity located at $x=0.5$.
Density and pressure in the two regions are given by $(\rho,\, p) = (1,\,1)$ for $x < 0.5$ and $(\rho,\,p) = (1/8,\, 1/10)$ for $x > 0.5$ while the velocity is zero everywhere.
We choose our physical units in such a way that the reference velocity is $v_0 = 5.25\times 10^5$ Km/s so that the temperature, in Kelvin, is expressed by $T = (p/\rho) v_0^2 \mu m_u/k_B$.
With this choice, the temperatures in the left and right regions become $T_L \sim 3852$ K and $T_R\sim 3345$ K.
The final solution at $t= 0.2$ is shown in Fig. \ref{fig:sodtest} for the adiabatic run (i.e. without explicit cooling).
The test is repeated also by including thermal losses and the solution is plotted in Fig. \ref{fig:sodtest_wc}.

\subsubsection{Adiabatic Shock Tube}
%

\begin{figure}
\centering
\includegraphics[width=1.0\columnwidth]{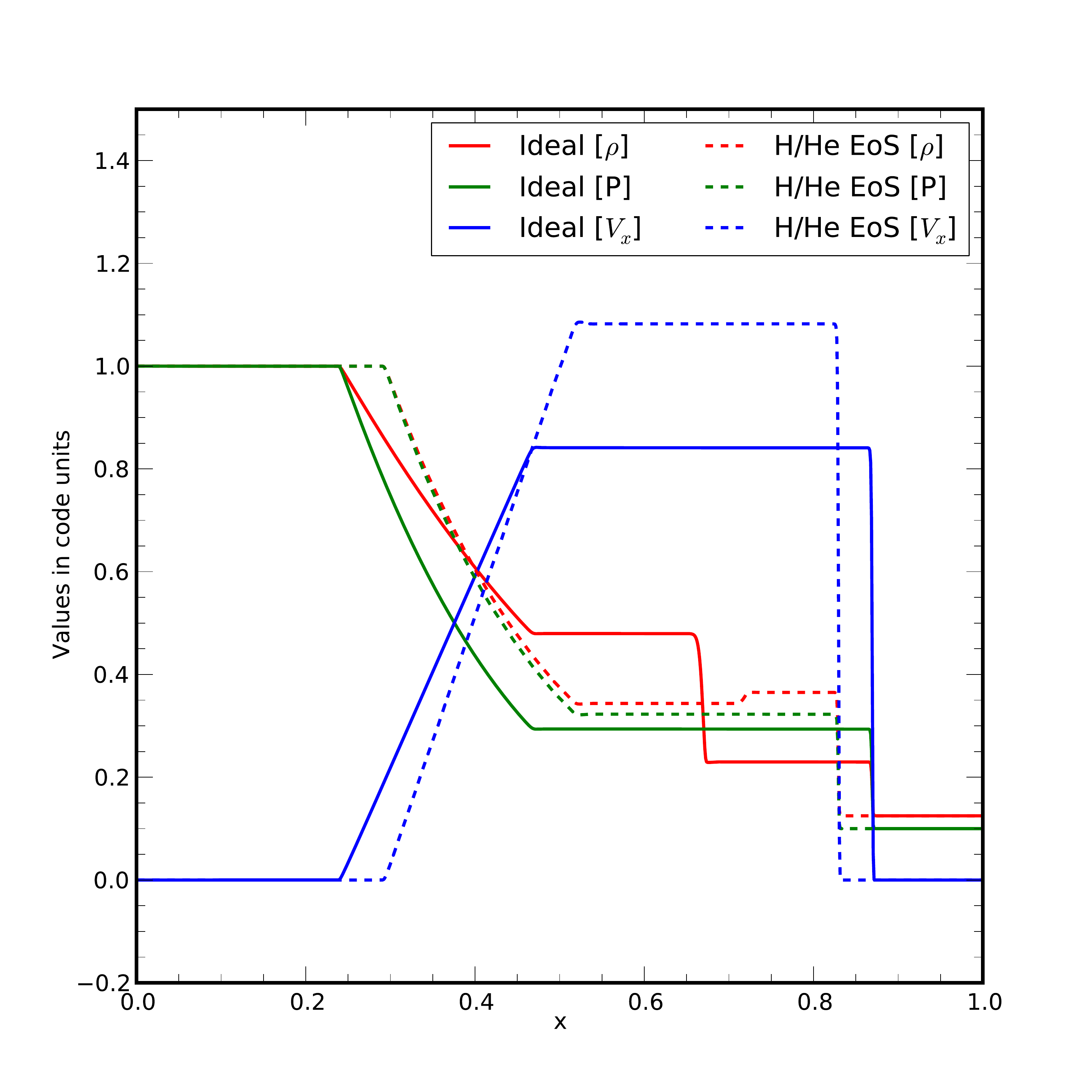}
\caption{\footnotesize The figure shows the variation of density $\rho$
  (\textit{red}), Pressure P (\textit{green}), and velocity $v_{x}$
  (\textit{blue}) along the X-axis (in code units)
  for a standard Sod Tube test (without explicit cooling) at time
  $\tau$ = 0.2. 
  The values obtained using an ideal EoS are shown as \textit{solid lines}
  while that obtained using a GammaLaw EoS are shown as \textit{dashed
    lines.} }
\label{fig:sodtest}
\end{figure}

Fig. \ref{fig:sodtest} compares the spatial variation of density, pressure and velocity for an ideal EoS (\textit{solid lines}) with that of a H/He EoS \textit{dashed lines}) without explicit cooling. 
The solution comprises, from left ro right, a rarefaction wave, contact discontinuity and a right-going shock.
Pressure and velocity are always continuous across the contact wave.
In the case of the gas with H/He EoS we obtain a larger compression ratio ($\approx 2.8$ compared to $\approx 1.84$ of the ideal gas) and the shock propagates slightly slower.
Similarly the tail and the head of the rarefaction wave both propagate at a smaller speed and this owes to a reduced value of the sound speed.
The density jump across the contact wave is largely reduced with the H/He EoS.

It is important to understand that in the case of an ideal EoS, only translation degrees of freedom contribute to the internal energy which is a linear function of the temperature, see Eq. (\ref{eq:ideal}).
In the case of the H/He EoS, however, the internal energy also takes contributions from additional degrees of freedom due to rotations and vibrations of di-atomic molecules (like H$_{2}$ at low temperatures).
These will not correspond to an increase in temperature.
Therefore, across a shock wave, the upstream kinetic energy will become available to the system not only to raise the temperature but also to populate the vibrational and rotational levels of the molecules (at lower temperatures), dissociate molecules and eventually ionize atoms (at larger temperatures).
Hence we expect the overall increase in temperature to be reduced in the case of H/He EoS as compared to that of an ideal EoS.

Finally we point out that while the employment of an ideal EoS gives a scale-free and self-similar solution to the Sod shock tube, the same argument does not hold for the H/He EoS which implicitly contains temperature scales corresponding to the above mentioned transitions.

\subsubsection{Radiative Shock Tube}
%

\begin{figure}
\centering
\includegraphics[width=1\columnwidth]{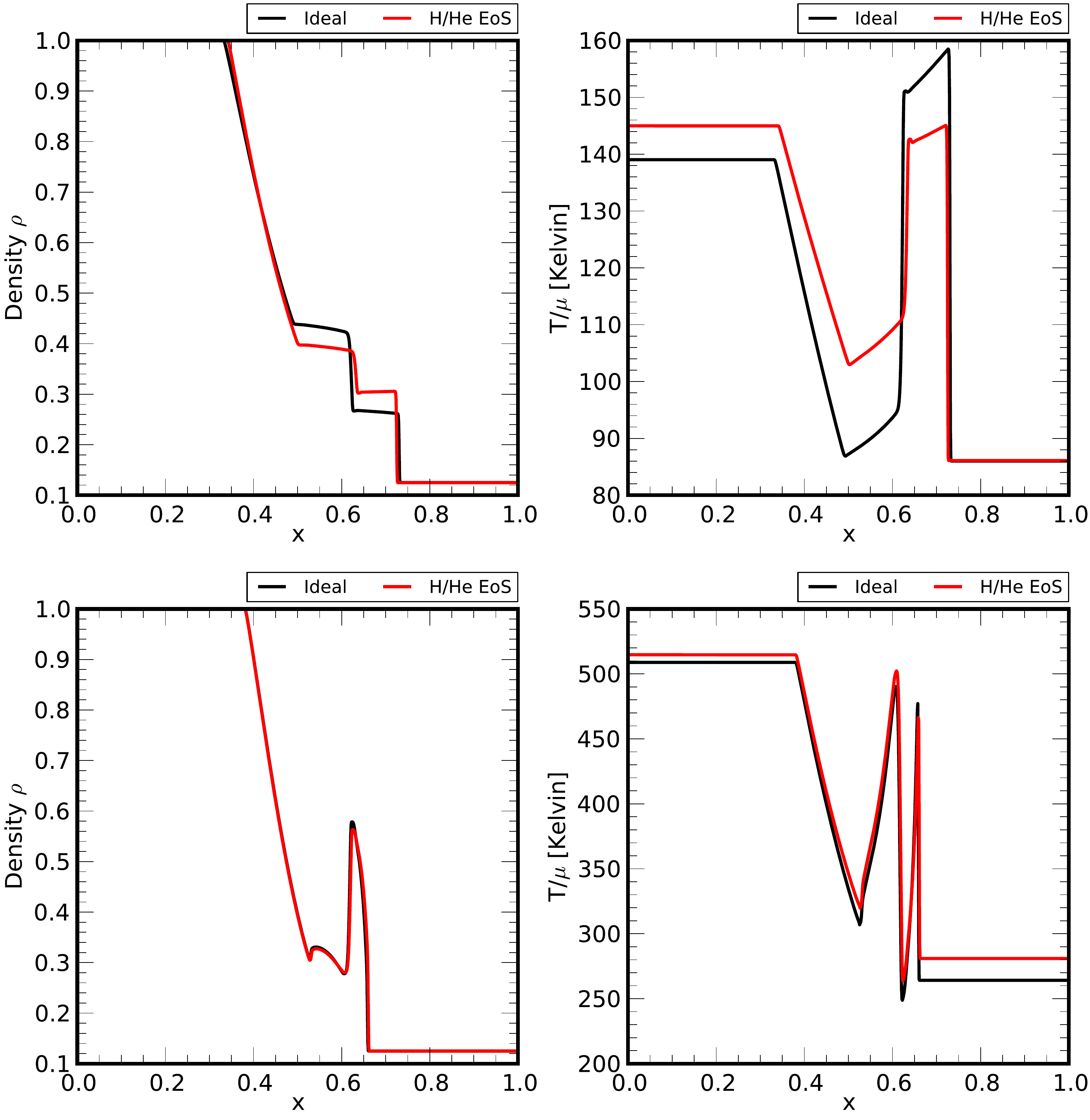}
\caption{\footnotesize The figure shows variation of density (\textit{left panels}) and
  temperature (\textit{right panels}) at the
  final stage of Sod shock tube test for cases which have included
  explicit cooling. The \textit{black line} represent values obtained
from ideal EoS, while those obtained using H/He EoS are shown in
\textit{red}. The top and bottom panels differ in the value of their
initial temperature on either sides of the interface at x = 0.5. }
\label{fig:sodtest_wc}
\end{figure}

We have repeated the Sod shock tube including the explicit non-equilibrium cooling described in Sect. \ref{sec:h2cool}.
We consider two initial conditions corresponding to different values of temperature.
In the first case the temperature of the left and right states is set to $T_L=400$ K and $T_R=200$ K, respectively.
The second case corresponds to hotter gas, $T_L = 4500$ K and $T_R =3500$ K.
Molecular and atomic hydrogen fraction are initially set to their equilibrium values at the local temperature (see Fig. \ref{fig:h2cooleq}).

Fig. \ref{fig:sodtest_wc} shows the density and temperature ($T/\mu$, $\mu$ being the mean molecular weight).
In all panels, solution obtained using an ideal EoS is shown as a \textit{black line}, whereas that obtained using H/He EoS with a \textit{red line}. 

For smaller initial temperatures (top panels), the density and temperature jumps differ by a modest factor ($\approx 20 \%$) and the positions of the waves is essentially the same.
The differences are not as large as in the adiabatic case since the H/He EoS, now given by Eq. (\ref{eq:HHe_cool}), only includes terms related to the internal degrees of freedom, see Sect. \ref{sec:h2cool}.
Terms related to dissociation and ionization correspond to energy losses and are thus accounted for by the cooling function $\Lambda$.
These terms are common to both EoS.
The presence of molecules at low temperature gives a considerable contribution to the internal energy budget.
Hence part of the thermal energy available at the shock serves in populating the internal molecular degrees of freedom rather than raising the temperature.
As a consequence, radiative losses are slightly less efficient for the H/He EoS.

For larger initial temperatures (bottom panels) differences are negligible and this behavior owes to a reduced cooling length scale. 
Indeed, in presence of a shock wave, the gas behind the front undergoes a rapid increase of the temperature while the level populations of $H_2$ do not occur instantaneously \citep{Flower2003}.
Since the cooling time scale is shorter than the average time scale over which level population takes place, internal energy changes are dominated by radiative losses.
Since the cooling function $\Lambda$ is the same for both EoS the final solutions are very similar.

These different test cases elucidates the importance of treating temperature-dependent EoS both in case when the system is in equilibrium and also when explicit radiative cooling is involved.

\subsection{MHD Shock Tube}
\label{ssec:sodtest}
%
%

Next we consider the collision between two magnetized fluids moving in opposite directions.
The problem is studied on the unit interval $x\in[0,1]$ with initial condition given as follows:
\begin{equation}
  \left(\rho,\, v_x,\, B_x,\, B_y,\, p\right) =
  \left\{\begin{array}{ll}
    \DS \left(5,\, 5,\, 5,\, 8,\, 10\right) & \mathrm{for}\quad
     x < 0.5 \\ \noalign{\medskip}
    \DS \left(1,\, -15,\, 5,\, 3,\,  1\right) & \mathrm{for}\quad
     x > 0.5
  \end{array}\right.
\end{equation}

The fluid on the left has larger inertia than the fluid on the right and it is also more magnetized.
We solve the problem by using linear interpolation on characteristic variables, the HLLD Riemann solver and an adaptive resolution consisting of a base grid of $128$ zones and $7$ levels of refinement with consecutive jump ratios of $2$ except for refining level $3$ where the grid jump is $4$.

\begin{figure*}
\centering
\includegraphics[width=2\columnwidth]{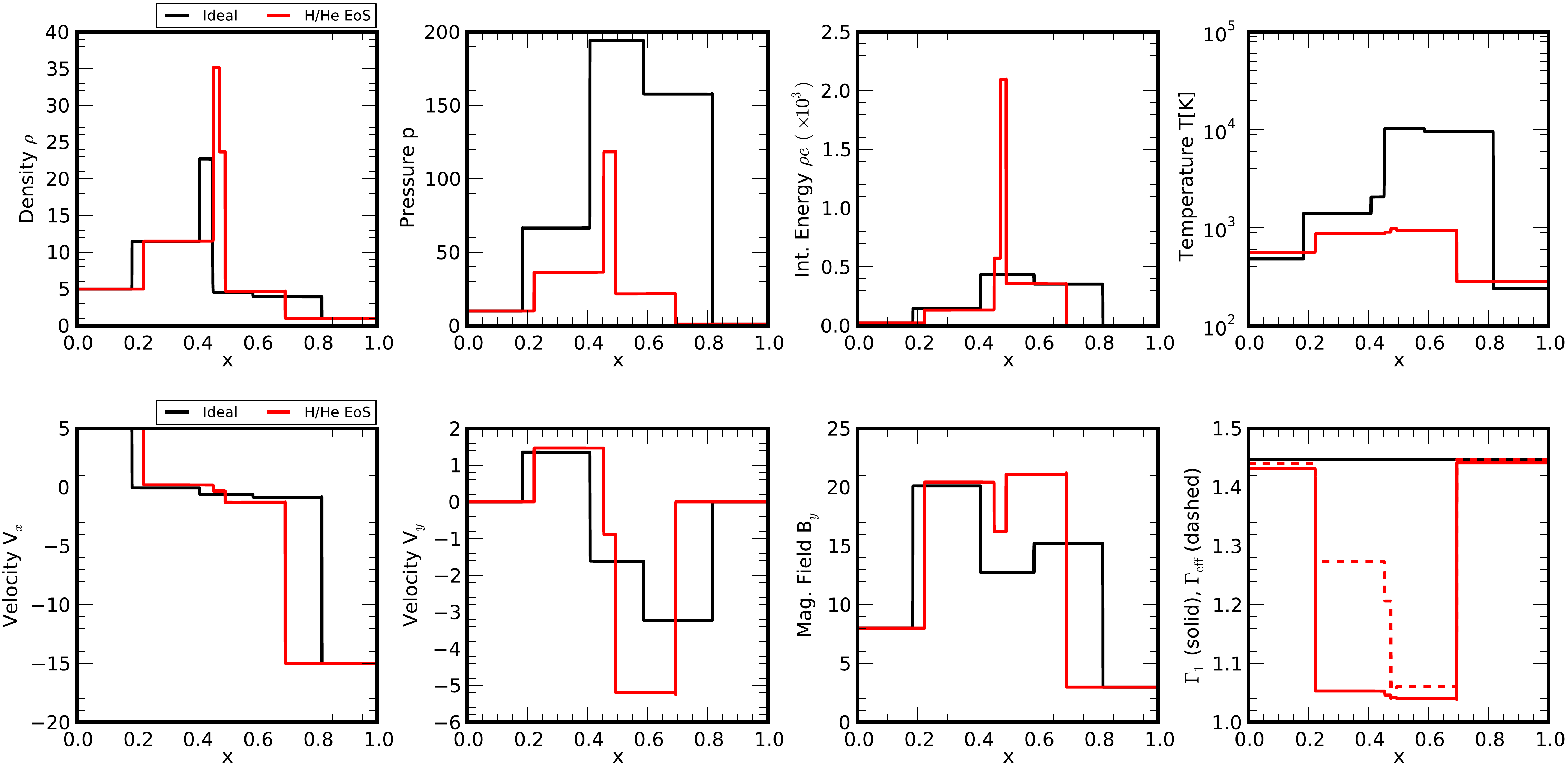}
\caption{\footnotesize The figure has 8 panels, going from \textit{top left} to
  \textit{bottom right} the panels show solution at time $t = 0.08$ of density, pressure, internal
  energy density ($\rho$ e), temperature, $X$ and $Y$ components of velocity,
  $Y$ component of magnetic field and adiabatic index respectively. 
  In each panel, quantities obtained
  for ideal EoS are shown in (\textit{black}) while the \textit{red}
  line denotes values obtained using H/He EoS.}
\label{fig:mhd_sod}
\end{figure*}

The solution is shown in Fig. \ref{fig:mhd_sod} at $t=0.08$ for the ideal and H/He EoS.
In order to carry out a fair comparison, we have chosen the adiabatic index of the ideal EoS to be $\Gamma = 1.447$ so that the gas internal energy has the same value for both EoS at $t=0$.
The wave pattern consists of a pair of outer fast magneto-sonic shocks enclosing two slow magneto-sonic shocks and a contact wave in the middle.

Overall, shocks are stronger and propagate faster for the ideal gas case leading to the formation of a more extended Riemann fan owing to a larger value of the sound speed (Eq. \ref{eq:cs2}).
This is evident by looking at the profiles of the first adiabatic index shown in the bottom right panel of Fig. \ref{fig:mhd_sod}.
For the H/He EoS, $\Gamma_1$ drops to $\approx 1.04$ in the shocked regions while it remains constant for the ideal EoS.
This effect becomes more pronounced at the two slow shocks which, in the case of H/He gas, propagate to the left at very small velocities forming a very thin shell of compressed material with large internal energy.

In spite of the reduced shock strength, however, the density compression attained at shocks is nearly the same or even larger for the H/He gas.
This behavior can be understood by noticing that lower values of the specific heat ratio should support larger density jumps. 
Indeed, the effective $\Gamma$-value, defined as $\Gamma_{\rm eff} = p/(\rho e) + 1$ (dashed line in the bottom right panel of Fig. \ref{fig:mhd_sod}), becomes smaller with the H/He EoS thus allowing for comparable or larger compression ratios (in density and magnetic field) even if the shock is weaker.

The EoS largely determines how the upstream kinetic energy is being converted into internal energy and heat: for the ideal gas an increase in internal energy corresponds to an increment in pressure and temperature while, in the range of temperature considered here ($10^2 \,{\rm K}\lesssim T \lesssim 10^4$ K), the same does not hold for the gas with H/He EoS.
In this case, in fact, the downstream internal energy is employed to excite internal vibrational and rotational levels without a corresponding increase in pressure or temperature.

\subsection{Blast Wave}
\label{ssec:blastwave}
%
%

\begin{figure*}
\includegraphics[width=2\columnwidth]{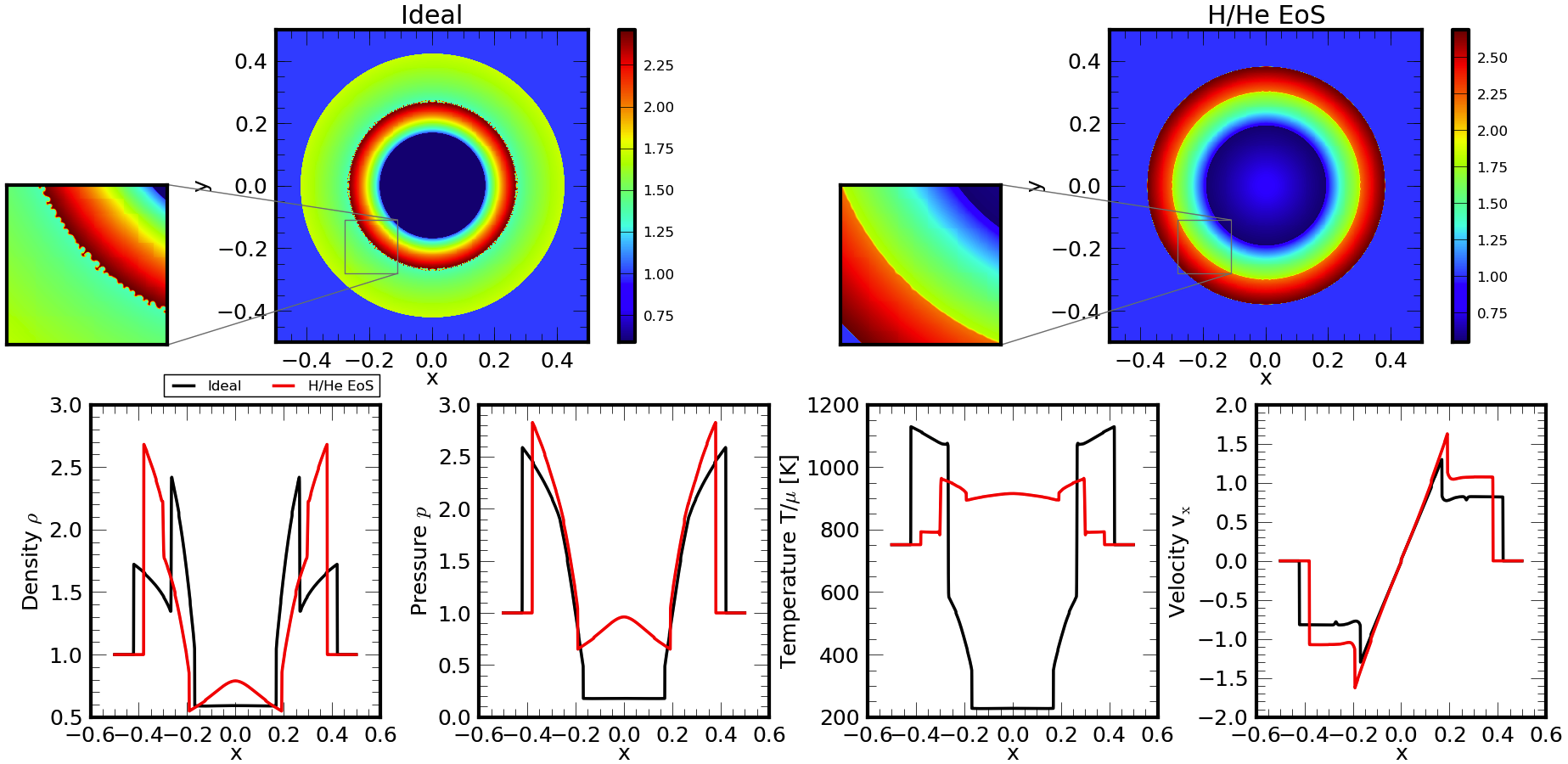}
\caption{\footnotesize \textit{Top panels:} Comparison of density $\rho$
  (in code units) for a hydrodynamic, spherical blast wave 
  at time $t$ = 0.15. 
  The \textit{left panel} shows the final shock structure 
  with an ideal EoS. The density obtained using the
  H/He EoS is shown in the \textit{top right panel}. 
  Alongside each panel a zoomed in view of density around the contact
  discontinuity is shown. The ideal EoS shows presence of
  Rayleigh-Taylor instability while its completely absent in case of
  H/He EoS. The \textit{bottom panels} from left to right compare 1D cuts at the
mid-plane of density $\rho$, pressure $p$, temperature (T/$\mu$) and
velocity (v$_{\rm x}$) for ideal EoS (\textit{black line}) and H/He EoS (\textit{red line}).}
\label{fig:blasttest}
\end{figure*}

We now consider a two-dimensional blast wave in Cartesian coordinates on the square domain  $-L_0/2 < x , y  < L_0/2$ where $L_0 = 2.5\times 10^{15}$ cm. 
An over-dense and over-pressurized region is initialized at the center of the domain ($x = y = 0$) inside a circular region of radius $r_0/L_0=0.1$ where density and pressure are set respectively to $\rho/\rho_a = 10$ and $p/p_a = 20$ in units of the corresponding ambient values, $\rho_a = n_a m_H$ and $p_a = \rho_a v_0^2$ (here we use $n_a = 10^5$ cm$^{-3}$ and $v_0 = 2.5$ km/s).
The computation is carried using adaptive mesh refinement (AMR) using a base grid of $256\times256$ grid zones and $4$ levels of refinement (effective resolution of $4096^2$ zones).
The MUSCL-Hancock with piecewise linear interpolation and the HLLC Riemann solver are employed.

The solution features an outermost forward shock wave followed by a contact discontinuity and a reverse shock as shown in the top panel of Fig. \ref{fig:blasttest} at $t = 0.15$.
In analogy with the Sod shock tube problem, we observe that the size of the spherical blast
wave is smaller when the H/He EoS is employed although the shock strengths (measured in terms of pressure jumps) are comparable for both EoS.
The compression ratio across the forward shock is noticeably larger for the H/He EoS ($\sim 2.7$) than for the ideal EoS ($\sim 1.7$) owing to a reduced value of the equivalent adiabatic index. 

An important difference lies in the structure of the contact wave as it can be noticed in the closeups in the bottom panel of Fig. \ref{fig:blasttest}.
In the ideal EoS case, the density contrast across the contact wave is favorable to the onset of the Rayleigh-Taylor instability while the same structure is stable for the H/He EoS since the density contrast is reversed (heavy fluid in front, light fluid behind).
Thus a shell of high density material is swept between the outermost shock and the contact wave.

From the computational perspective, we have compared the ideal and H/He EoS in terms of speed.
Our results show that the average wall clock time per numerical step is $0.05$ s with the ideal EoS while we found $0.25$ s and $0.9$ s for the tabulated and the root finder approaches, respectively, in the case of H/He EoS. 
As expected the computation with constant $\Gamma$ are the fastest ones.
However, it is interesting to note that pre-computed tabulated values give a speed up of about $4$ times than that obtained using the root finder approach, while maintaining essentially the same accuracy in the final solution.

\subsection{One-dimensional pulsed molecular jets}
%
%

As an astrophysical applications, we consider in the next test the propagation of velocity pulsations in a 1D molecular jet model that includes radiative losses.
Multidimensional extensions of this study will be considered in forthcoming papers.

In order to resolve the thin post-shocked regions we employ adaptive mesh refinement to enhance resolution in high-temperature and dense regions.
The computational domain extends from $z=0$ up to $z=32$ in units of the jet radius $r_j \sim 75$ AU and it is covered by a base grid of $256$ grid zones.
We employ $10$ levels of refinement yielding an equivalent resolution of $262,144$ zones.
The domain is initially filled with a fully molecular ($X_{H_2} \approx$ 0.5) ambient representing a young star-forming core at the temperature of $T_a = 50$ K and decreasing density so that  
\begin{equation}
  \rho_a(z) = \frac{\rho_a}{1 + (z/z_0)^2}\,,\qquad
  p_a(z)    = \frac{\rho k_BT_a}{\mu m_H} \,.
\end{equation}
In the previous equation $\rho_a$ is the ambient density at $z=0$ while $z_0 = 10 r_j$.
An over-dense jet ($\rho_j/\rho_a = 3$) is injected from the nozzle at $z = 0$ with temperature $T_j = 10^3$ K resulting in an over-pressurized jet ($p_j/p_a \approx 60$).
The initial hydrogen fractions at the nozzle are estimated based on equilibrium shown in
Fig. \ref{fig:h2cooleq} while the jet density is assumed to be $\rho_j = n_j m_H$ with $n_j = 10^4$ cm$^{-3}$.
The injection velocity has sinusoidal perturbations of the form $v_j=v_0\,(1 + 0.25\sin(2\pi\, t/T_p$)), with base velocity  $v_0= 70$ km s$^{-1}$ and pulsation period $T_p = 40$ years. 

As the system evolves, velocity pulsations steepen into a chain of forward/reverse shock pairs, the first of which becomes the strongest one reaching temperatures of $T\sim 10^5$ K where hydrogen is quickly ionized.
Under these physical conditions the two EoS yields essentially the same results since thermal kinetic motion is characterized only by translational degrees of freedom.

\begin{figure*}
\includegraphics[width=2\columnwidth]{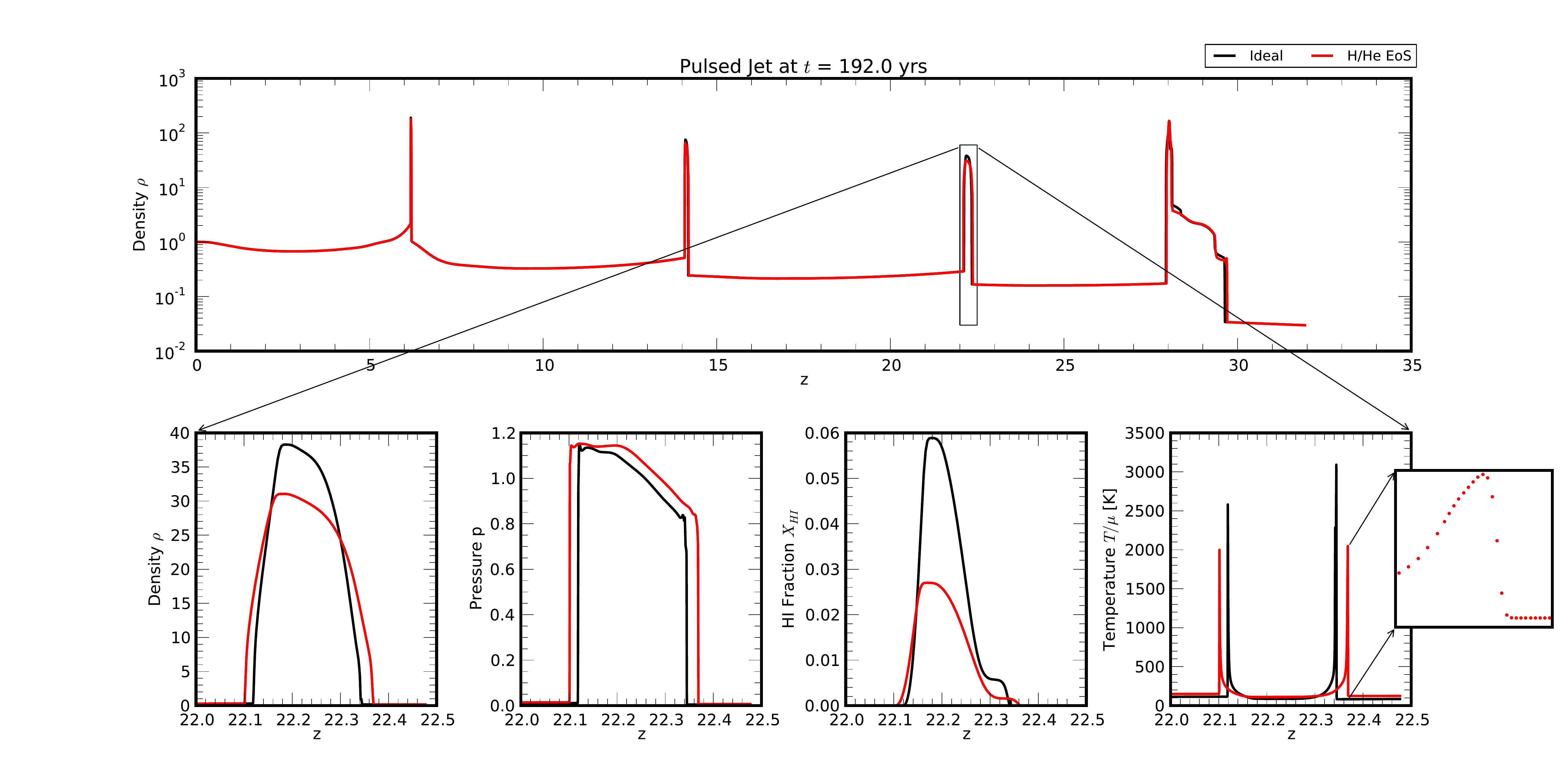}
\caption{\footnotesize Results from the test studying the propagation of supersonic
  pulsations in molecular medium. \textit{Top panel} shows the
  comparison of density (in code units) for ideal EoS (\textit{black
    line}) and H/He EoS (\textit{red line}) on a log scale at time $t =
  192$ yrs. The \textit{bottom panels} compares zoomed up quantities for a particular knot marked
  within the rectangle in the top panel using the same color keys. 
  These quantities from left to right are density $\rho$, pressure
  $p$, HI fraction $X_{HI}$ and temperature T/$\mu$. A zoomed inset of
  the forward shock obtained from H/He EoS is shown as well with
  \textit{red dots} to demonstrate the importance of resolving the
  shock (see text).  
  }
\label{fig:puljettest}
\end{figure*}

Successive pulses, however, form weaker shocks with temperatures of few thousand K, see Fig. \ref{fig:puljettest}.
As the cooling time is much smaller than the dynamical time, the material behind the shocks start to cool radiatively in a very efficient way.
Temperature peaks immediately downstream of the forward and reverse shocks and quickly drops afterwards forming extremely thin layers of hot material (see the enlargement in the right panel of Fig. \ref{fig:puljettest}).
The layer of fluid between the forward/reverse shock pairs becomes nearly isobaric with a high density pile. 
We point out that the numerical resolution of such layers is crucial in order to correctly capture the physical processes (such as ionization and radiative losses) taking place on different time scales.
The importance of grid resolution has also been addressed by \cite{Tesileanu2009} in the context of time-dependent radiative shocks subject to atomic cooling.

As it has been elucidated in earlier tests, the transformation of upstream kinetic energy into internal energy depends on the EoS considered.
While pressure is essentially the same for both EoS, the temperature values reached immediately behind the shocks are larger for the ideal EoS gas than for the H-He EoS because part of the internal energy is employed to dissociate molecules.
As a consequence the amount of atomic hydrogen produced during the dissociation process is halved ($\sim 2.7\%$ for the H-He EoS vs $\sim 5.9 \%$ with the ideal EoS).

\subsection{Axisymmetric MHD jets}
%
%

\begin{figure*}
\includegraphics[width=2\columnwidth]{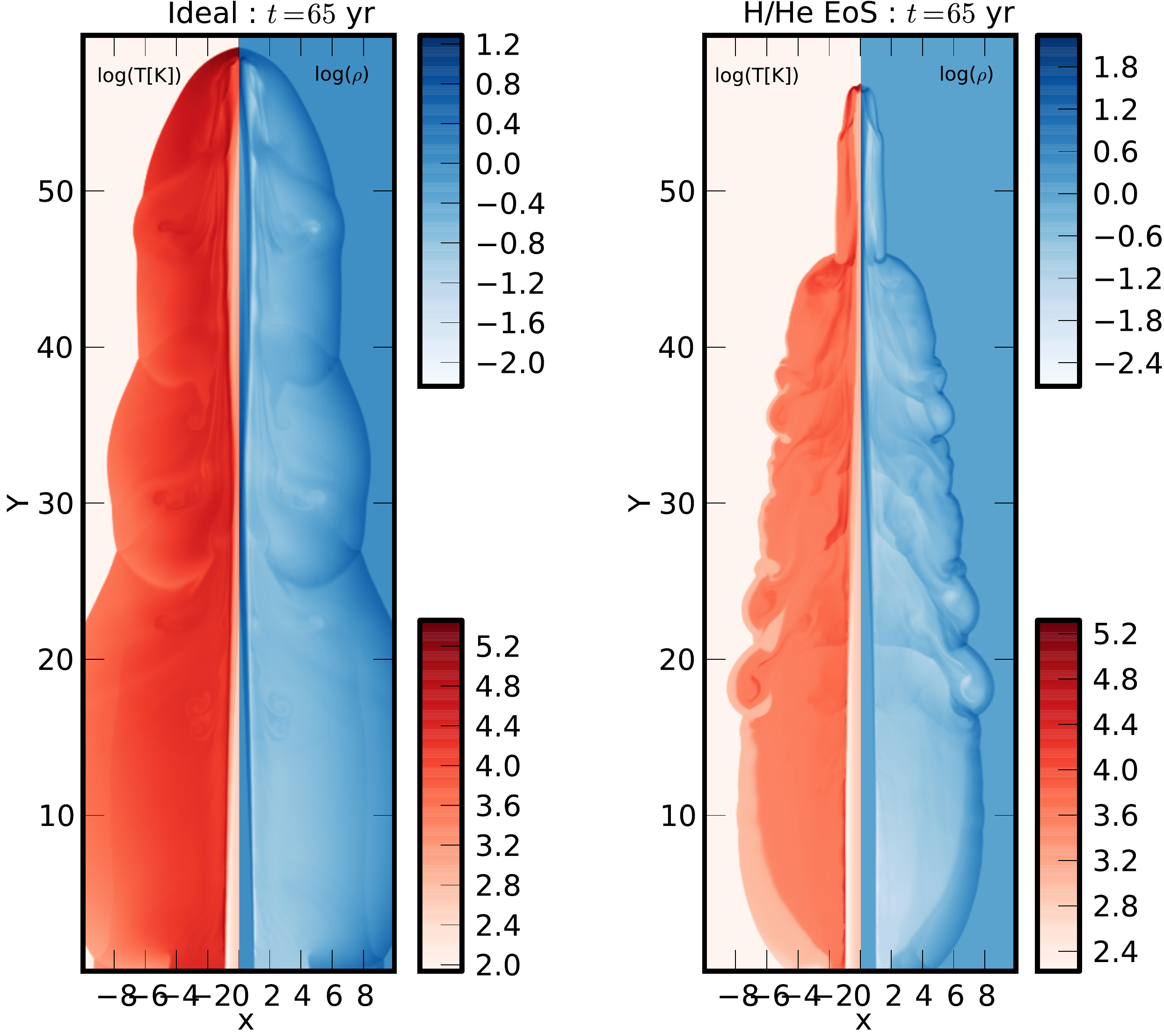}
\caption{\footnotesize Comparison of logarithmic values both of density, $\rho$
  (right half in {\it{blues}}) and temperature in Kelvin (left half in {\it{red}}) for a 2D axisymmetric jet without
  explicit cooling. The \textit{left panel} shows the jet structure 
  with an ideal EoS, while, that obtained from a H/He EoS
  is shown in the \textit{right panel}.}
\label{fig:jet2dtest}
\end{figure*}

In this test problem we consider the effect of the H/He EoS on the propagation of an axisymmetric MHD jet using 2D cylindrical coordinates.
For simplicity, we consider the MHD equations without the radiative loss term.
The computational domain is defined by $r\in[0,10]$ and $z\in[0,60]$ and discretized using $256\times 1536$ grid zones.
It is initially filled with a static ambient medium with constant density $\rho_a = m_Hn_a$ ($n_a = 10^3$ cm$^{-3}$) and pressure $p_a=(3/5) \rho_av_0^2$ where $v_0 = 1$ km/s is our unit velocity.
The ambient medium is threaded by a constant vertical magnetic field $B_z = \sqrt{2\sigma_z p_a}$ where $\sigma_z=1/4$ is the magnetization parameter corresponding to the vertical field component.

A supersonic inflow is imposed through a circular nozzle at the lower boundary ($z=0$) with radius  $R_j = 20$ AU where the beam is inject with the same density as the ambient medium and is injected with a constant vertical velocity $v_z = 150$ km s$^{-1}$.
The jet carries an azimuthal magnetic field with strength given by \textbf{(\citealt{Tesileanu2008})}
\begin{equation}
  B_\phi(R) = \left\{\begin{array}{ll}
    \DS B_m\frac{R}{a} & \quad\mathrm{for}\quad R < a  \\ \noalign{\medskip}
    \DS B_m\frac{a}{R} & \quad\mathrm{otherwise}  \,,
  \end{array}\right.
\end{equation}
where $B_m = \sqrt{4\sigma_\phi p_a/(a^2(1 - 4\ln a))}$, $\sigma_\phi=1$ controls the strength of the azimuthal magnetic field and $a=0.8$ is the magnetization radius.
Radial balance across the jet beam leads to the following (thermal) pressure profile:
\begin{equation}
  p(R) = p_a + B_m^2\left[1 - \min\left(\frac{R^2}{a^2},1\right)\right] \,,
\end{equation}
where the integration constant is chosen in such a way to have pressure balance across the jet border.

As the jet enters into the ambient medium, it immediately forms a bow-shock that pushes the ambient material to its sides.
The overall morphology of the jet is determined primarily by the Mach number, the density contrast (a result that has also been confirmed in laboratory experiments \citep[][]{Belan2013, Belan2014}) and the magnetic field strength.
The processed material gets heated and forms the cocoon as shown in Fig. \ref{fig:jet2dtest} at $t \sim 65$ yrs for the ideal (\textit{left}) and H/He EoS (\textit{right)}. 
The left half of each panel shows the logarithmic values of temperature in Kelvin, while the right panel shows the corresponding value of density in units of the ambient density.
For both EoS considered here, the largest temperature is achieved at the bow-shock where $T_{\max} \approx 2.9\times 10^5$ K for the ideal EoS while $T_{\max}\approx 2\times 10^5$ K for the H/He EoS.
On the contrary, the maximum density obtained with H/He EoS is $n_{\max} \approx 1.8\times10^{5}$ cm$^{-3}$, larger than its ideal counterpart by a factor of $\approx 10$.

The two simulations differ mostly in the overall morphology.
While the ideal EoS gas tends to produce a larger cocoon, the H/He gas results in a thinner and colder cocoon, a structure that resembles that of a radiative jet.
Once again, we notice that the shocked ambient medium contributes more to raising the internal energy rather than the temperature inside the cocoon.
This results in a lowered thermal pressure to support the material feeding the cocoon leading to the formation of a thin dense layer on its sides.
This thin layer also shows formation of rolled up Kelvin-Helmholtz vortices.

Similar to the blast wave test (see \S \ref{ssec:blastwave}), we have performed a computational efficiency comparison.
We found the average wall clock time per step  to be $0.048$ s for the constant-$\Gamma$ EoS while, for the H/He EoS, the timing were $0.095$ s using tabulated approach and three times as much using the root-finder conversion method ($\Delta t_{\rm av} = 0.29$ s).
These results confirm the trend already established for previous tests.

\section{Summary}
\label{sec:summary}
%
%
%
%

In this work, we have revisited fundamental thermodynamic properties for the modeling a gas mixture in terms of its thermal and caloric equations of state.
This has been achieved by consistently including temperature-dependent physical processes such as ionization, dissociation and level population of internal degrees of freedom.
This approach has shown to considerably improve over the (widespreadly used) constant-$\Gamma$ EoS in those astrophysical environments where additional physical processes play a vital role. 
In this respect, we consider the H-He EoS that takes into account various atomic and molecular processes in the context of equilibrium conditions as well as in the presence of non-equilibrium cooling.

The paper also presents  a detailed numerical implementation of thermally ideal and general caloric convex EoS for Godunov-type numerical schemes.
In particular, it has been shown that conservative schemes require two major modifications.
The first one concerns the Riemann solver and only Jacobian-free solvers have been considered in the present context.
For such schemes only a change to the sound speed is necessary thus making their extension to a more complex (convex) EoS straightforward.
The second modification relates to the conversion between total conserved energy and gas pressure.
Since a general thermal EoS can now be nonlinear functions of the temperature, a closed-form expression between internal energy and pressure cannot be obtained and the problem must be treated numerically.
To cope with this, we have explored two alternative strategies: the first one relies on numerical inversion through root-finder methods whereas the second one is based on a combination of lookup table and interpolation.
We found the first approach to be accurate but also computationally more expensive (by a factor $\sim 20$) when compared to the Ideal EoS.
The second approach, best suited for equilibrium conditions or tabulated EoS, yields similar accuracy and it largely reduces the computational workload (a factor $3\sim 4$ in the presented benchmarks).
Cubic spline interpolation (for the tabulated approach) is preferred over bilinear interpolation in order to ensure thermodynamic consistency so as to avoid the loss of convexity and the ensuing formation of composite waves in the solution of the Riemann problem.

The proposed numerical benchmarks indicate that the differences between the ideal and H-He gas can be considerable when the gas temperature lies in a range favorable to dissociation or ionization processes as well as to populate levels corresponding to roto-vibrational degrees of freedom.
The differences are particularly evident in problems involving shock propagation.
For a mono-atomic ideal EoS (constant $\Gamma$), in conditions of local thermodynamic equilibrium, the energy generated during the shock impact can only be distributed into  translational degrees of freedom thus corresponding to a temperature increment.
Conversely, for the H-He EoS, since additional degrees of freedom are available, the internal energy of the post-shock gas becomes available for populating rotational and vibrational levels rather than resulting in an increase in temperature.
In addition, the employment of such an EoS for temperatures $10^3\lesssim T \lesssim 10^4$ (depending on the density) results in an lower effective adiabatic index and larger compression ratios are observed at shocks.

These features significantly alter the structure of the solution of the Riemann problem in terms of wave strengths and positions and frequently result in the formation of thin-shell of high density gas.
In a 2D environment, for example, the density contrast across the contact wave may even reverse when switching from the ideal to the H-He EoS thereby completely quenching the growth of the Rayleigh-Taylor instability (see the blast wave test).

Under non-equilibrium conditions when optically thin cooling is introduced, a similar trend is observed although differences between the EoS are visible mainly at lower temperature range ($T \sim 100$ K).
In such a case, terms related to the molecular degrees of freedom are added to the internal energy while the terms responsible for radiative cooling are accounted for by the cooling function.
We stress out that very high spatial resolution may be required to resolve the post-shock regions since the cooling length becomes much smaller than the overall spatial scale. 
On the contrary, in the high temperature regime where hydrogen is mostly ionized, the behaviors of the ideal and H-He EoS become similar as the expressions for the internal energy (and the radiative losses) are essentially for the same.

The numerical methods described here have been implemented in the PLUTO code \citep{Mignone2007} starting with version 4.2 and are now publicly available.
We believe that this set of tools can become useful in the investigation of various astrophysical systems like planet formation, molecular jets from young stars, gravitational collapse of molecular cloud, galaxy formation due to cold flows etc.

\begin{acknowledgements}
We would like to thank the referee, Prof. A. Raga, for his constructive
suggestions on the paper. BV is grateful to the funding support from the University of Torino 
under the contract: "Progetto di Ateneo-Compagnia di SanPaolo". 
\end{acknowledgements}

\bibliographystyle{aa}

\begin{appendix}
\section{Fundamental Gas Derivative}
\label{sec:G}
%
%
%

Using standard thermodynamic relations, equation (\ref{eq:G}) can be expressed as a function of $\rho$ and temperature  $T$ \citep{Nannan2013259}
\begin{equation} \label{eq:G2}
  \mathcal{G} = \frac{1}{2c_s^{2}\rho^{3}}(\Gamma_1 + \Gamma_2 + \Gamma_3)
\end{equation}
where the thermodynamic speed of sound, $c_s$, is computed from
\begin{equation}
  c_s^{2} =   \left(\frac{\partial P}{\partial \rho}\right)_T
            + \frac{T}{C_v\rho^2}\left(\frac{\partial P}{\partial T}\right)_\rho^{2}
\end{equation}
and the three $\Gamma$'s are given by,
\begin{equation}
  \begin{array}{lcl}
    \Gamma_1 &=& \DS \rho^{4} \left(
          \frac{\partial^{2} P}{\partial\rho^{2}}\right)_T
      +  2\rho^{3} \left(\frac{\partial P}{\partial \rho}\right)_T
    \\  \noalign{\medskip}
    \Gamma_2 &=& \DS 3\frac{T\rho^{2}}{C_{v}}
                \left(\frac{\partial P}{\partial T}\right)_\rho
                \left(\frac{\partial^{2} P}{\partial \rho \partial
    T}\right)_{T,\rho}
    \\  \noalign{\medskip}
    \Gamma_3 &=& \DS \left[\frac{T}{C_v}\left(\frac{\partial P}{\partial
    T}\right)_\rho\right]^{2}\times\left[3 \left(\frac{\partial^{2}
  P}{\partial T^{2}}\right)_\rho + \frac{1}{T}\left(\frac{\partial
  P}{\partial T}\right)_\rho\left(1 - \frac{T}{C_v}
  \left(\frac{\partial C_v}{\partial T}\right)_\rho\right)\right]
 \end{array}
\end{equation}
where, C$_{v}$ denotes specific heat capacity at constant volume.
%

\section{Spurious formation of composite waves}
\label{sec:Gtest}
%
%
%
It has been shown in Section \ref{sec:table_conversion} that lookup table methods can be faster than nonlinear root-finder methods (Sect. \ref{sec:root_conversion}) although care must be taken to ensure consistency with thermodynamic principles.
Indeed, as stated in Sect. \ref{sec:interpolation}, inaccurate interpolation may give rise to a local loss of convexity with the ensuing formation of composite waves in the solution of the Riemann problem.
This failure can be ascribed to the choice of a low-order interpolant and can be quantified by  numerical computation of the fundamental gas derivative, Eq. (\ref{eq:G2}).

\begin{figure*}
\includegraphics[width=2\columnwidth]{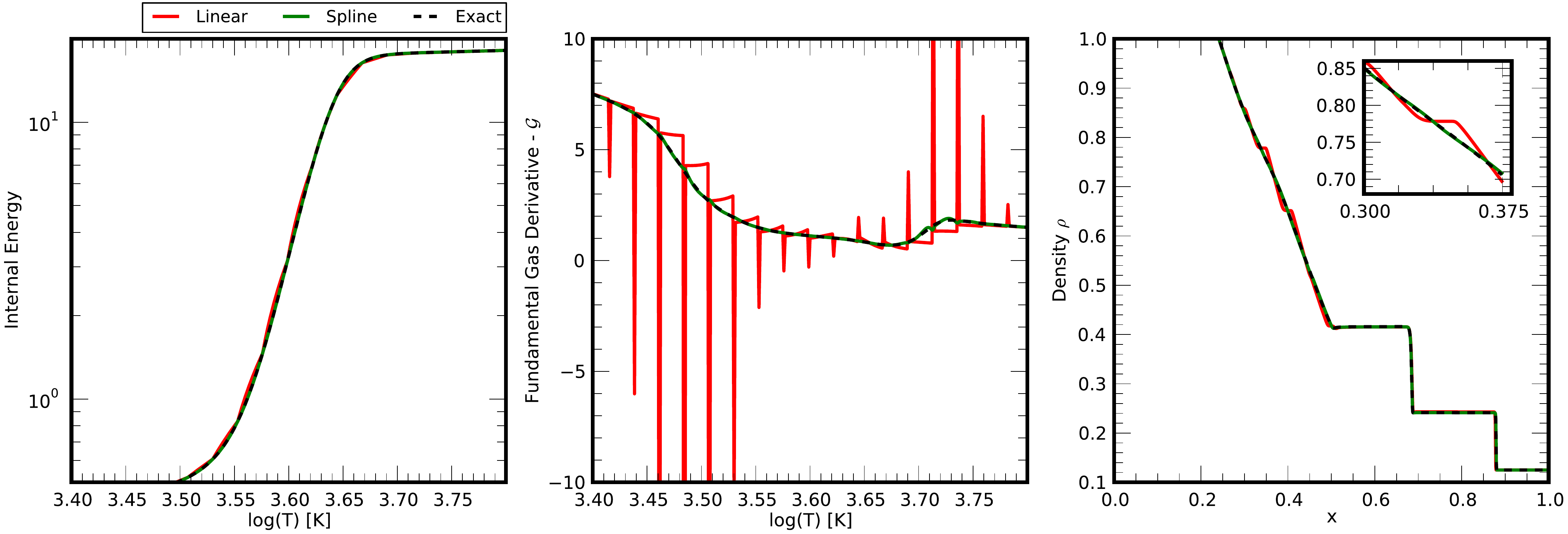}
\caption{\footnotesize Comparisons between exact and tabulated approaches for
the partially hydrogen gas EoS.
The black dashed line corresponds to computations obtained with the iterative
root-finder approach outlined in Section \ref{sec:root_conversion} while
the red and green solid lines corresponds to the tabulated approach presented
in Section \ref{sec:table_conversion}.
\textit{Left}: gas internal energy (Eq. \ref{eq:appinte}) around the transition region.
\textit{Middle}: fundamental gas derivative $\mathcal{G}$ (Eq. \ref{eq:G2})
computed using standard central differencing.
The large spikes in the red curve are due to discontinuous first derivatives
using linear interpolation between adjacent node values in the table.
\textit{Right}: Density profile in the Sod shock tube at $t=0.2$ using the exact
and spline approaches. It is clear the appearance of spurious compound structures in
the rarefaction wave when a linear spline is used to approximate the table. 
}
\label{fig:appfig1}
\end{figure*} 

Here we present the analysis for a partially ionized hydrogen gas in LTE equilibrium with internal energy given by
\begin{equation}\label{eq:appinte}
 e = \frac{3}{2} k_{b} T + \chi
\end{equation}
where $\chi$ is the ionization energy for hydrogen.

The internal energy is characterized by a rapid increase around $T\approx 4000$ K (corresponding to ionization) and it is plotted in the left panel of Fig. \ref{fig:appfig1} together with piecewise polynomial approximations using linear (Eq. \ref{eq:linear_spline}, red curve) and cubic (Eq. \ref{eq:cubic_spline}, green curve) splines.
The table was constructed by sampling the temperature range $[1,10^8]$ K with $N_c=350$ points using a uniform log spacing $\Delta\log T = 8/350.$
Clearly, the transition region is poorly sampled and the linear interpolation manifestly reveal that the first derivative is not continuous.
This behavior has dramatic consequences when plotting the profile of the fundamental derivative computed using central differences and shown in the middle panel of Fig. \ref{fig:appfig1}.
Large spikes are evident at the juncture points between contiguous piecewise linear polynomials while this effect is not seen using monotone spline interpolation.
The large over- and under-shoots fail to preserve local convexity leading to negative values of $\mathcal{G}$.
When applied to a shock-tube test problem, this loss of convexity results into composite waves as observed in the rarefaction branch shown in rightmost panel of the figure.
The number of spikes  can be reduced by increasing the number of sample points for linear
interpolation to $512$ although it is completely absent when using cubic spline interpolation.

\section{Monotone Spline}
\label{sec:monotone_spline}
%
%
%

Given a set of values $f_i=f(T_i)$ and theirs derivatives $f'_i=df/dx(T_i)$ defined at node points $T_i$ we construct a piecewise cubic approximation on each interval $[T_i,\,T_{i+1}]$ in the form
\begin{equation}\label{eq:cubic}
  f_i(T) =   a_i \left(\frac{T-T_i}{\Delta T_i}\right)^3
           + b_i \left(\frac{T-T_i}{\Delta T_i}\right)^2
           + c_i \left(\frac{T-T_i}{\Delta T_i}\right) + d_i
\end{equation}
where $\Delta T_i = T_{i+1}-T_i$ and, for simplicity, we have dropped the index $j$ spanning along the density grid. 
Cubic polynomials are frequently used as they allow to smoothly fit a set of data values with specified derivatives at each endpoint.
Although continuity of the first and second derivatives across all polynomial segments are normally invoked when deriving the coefficients of Eq. (\ref{eq:cubic}), we shall pursue a different approach where ${\cal C}^2$ continuity is sacrificed in favor of monotonicity \citep{WolbergAlfy2002}.
The resulting spline will be the smoothest curve that passes through the control points while preserving monotonicity of the dataset but not necessarily continuity of the second derivative.
It can then be shown that the coefficients are  
\begin{equation}
  \begin{array}{lcl}
  a_i &=& \DS
   \Delta T_i\left(-2m_i + f'_i + f'_{i+1}\right)
  \\ \noalign{\medskip}
  b_i &=&\DS
   \Delta T_i\left(3m_i - 2f'_{i} - f'_{i+1}\right)
  \\ \noalign{\medskip}
  c_i &=& \Delta T_if'_i
  \\ \noalign{\medskip}
  d_i &=& f_i
  \end{array}
\end{equation}
where
\begin{equation}
  m_i = \frac{f_{i+1} - f_i}{T_{i+1} - T_i} \,.
\end{equation}

The cubic is monotone if there is no change in sign in the interval $[x_i,x_{i+1}]$.
If this holds, then a necessary conditions is that
\begin{equation}\label{eq:spline_sgn}
  {\rm sgn}(f'_i) = {\rm sgn}(f'_{i+1}) = {\rm sgn}(m_i) 
\end{equation}
Defining $\alpha_i = f'_i/m_i$ and $\beta_i = f'_{i+1}/m_i$, and assuming that $m_i\neq 0$, the sufficient conditions for monotonicity can be summarized as follows:
\begin{enumerate}
  \item Eq. (\ref{eq:spline_sgn}) \emph{and} $\alpha_i + \beta_i - 2 \le 0$;
  \item Eq. (\ref{eq:spline_sgn}) \emph{and} $\alpha_i + 2\beta_i - 2 > 0$
        \emph{and} one of the following conditions:
    \begin{itemize}
      \item[a)] $2\alpha_i + \beta_i - 3 \le 0 $;
      \item[b)] $\alpha_i + 2\beta_i - 3 \le 0 $;
      \item[c)] $\alpha_i^2 + \alpha_i(\beta_i-6) + (\beta_i-3)^2 < 0$.
    \end{itemize}
\end{enumerate}
If monotonicity cannot be satisfied inside the interval or the spline does not have enough curvature, we simply revert to linear interpolation.
A test for the latter condition is simply $|d^2f_i(T)/dT^2| < \epsilon|f_i(T)|$ computed at the interval midpoint $T = (T_i+T_{i+1})/2$ and $\epsilon=10^{-16}$ is a small number.
\end{appendix}

\end{document}